\documentclass[pdflatex,sn-basic]{sn-jnl}

\jyear{2021}%

\theoremstyle{thmstyleone}%
\newtheorem{theorem}{Theorem}
\newtheorem{proposition}[theorem]{Proposition}%
\newtheorem{corollary}[theorem]{Corollary}
\newtheorem{lemma}{Lemma}%
\usepackage{amsmath,bm}
\theoremstyle{thmstyletwo}%

\theoremstyle{thmstylethree}%
\usepackage{color}
\usepackage{ulem}
\usepackage{subfigure}
\usepackage{optidef}

\usepackage{multicol}
\usepackage{bbm}

\usepackage{siunitx}
\begin{document}

\title[Derivatives-based portfolio decisions]{Derivatives-based portfolio decisions. An expected utility insight}

 
\author[1]{\fnm{Marcos} \sur{Escobar-Anel}}\email{marcos.escobar@uwo.ca}
\equalcont{These authors contributed equally to this work.}
\author[1,2]{\fnm{Matt} \sur{Davison}}\email{mdavison@uwo.ca}
\equalcont{These authors contributed equally to this work.}

\author*[1]{\fnm{Yichen} \sur{Zhu}}\email{yzhu562@uwo.ca}
\equalcont{These authors contributed equally to this work.}

\affil*[1]{\orgdiv{Department of Statistical and Actuarial Sciences}, \orgname{Western University}, \orgaddress{\street{1151 Richmond Street}, \city{London}, \postcode{N6A 5B7}, \state{Ontario}, \country{Canada}}}

\affil[2]{\orgdiv{Department of Mathematics}, \orgname{Western University}, \orgaddress{\street{1151 Richmond Street}, \city{London}, \postcode{N6A 5B7}, \state{Ontario}, \country{Canada}}}


\abstract{This paper challenges the use of stocks in portfolio construction, instead we demonstrate that Asian derivatives, straddles, or baskets could be more convenient substitutes.  Our results are obtained under the assumptions of the Black--Scholes--Merton setting,  uncovering a hidden benefit of derivatives that complements their well-known gains for hedging, risk management, and to increase utility in market incompleteness. The new insights are also transferable to more advanced stochastic settings.\\     
The analysis relies on the infinite number of optimal choices of derivatives for a maximized expected utility (EUT) agent; we propose risk exposure minimization as an additional optimization criterion inspired by regulations. Working with two assets, for simplicity, we demonstrate that only two derivatives are needed to maximize utility while minimizing risky exposure. In a comparison among one-asset options, e.g. American, European, Asian, Calls and Puts, we demonstrate that the deepest out-of-the-money Asian products available are the best choices to minimize exposure. We also explore optimal selections among straddles, which are better practical choice than out-of-the-money Calls and Puts due to liquidity and rebalancing needs. The optimality of multi-asset derivatives is also considered, establishing that a basket option could be a better choice than one-asset Asian call/put in many realistic situations.\\
}



\maketitle

\section{Introduction}


Financial derivatives such as futures, options and swaps play essential roles in current financial markets. They are used for hedging and speculation as well as for arbitrage opportunities. The history of derivatives is as old as the history of commerce. Derivatives have grown into an indispensable asset class since the 1970s, in part due to the increase in volatility and complexity of global financial markets. The popularity of derivatives is such that some market analysts, such as \cite{maverick_2020}, place the size of derivatives at more than 10 times the total world gross domestic product (GDP). Vanilla European and American options usually come to mind first when discussing options. Beyond that, a wide variety of options are traded in centralized exchanges or over-the-counter (OTC) markets, and some investors are even able to define their own products and terms. The enduring appeal of derivatives lies in their diversity and hence the capacity to fulfill their needs of financial players.


This paper uncovers an additional benefit of derivatives. It addresses a basic, yet poorly understood question for investors: what is the best financial derivative to include in a portfolio? We challenge the common practice of using the underlying stock; instead, we demonstrate that Asian derivatives, straddles or baskets could be more convenient choices. Our analysis and results are obtained under the safe assumptions of a Black--Scholes--Merton model, which not only uncovers the hidden  ``in plain sight'' benefits of derivatives, but also highlights their potential for applications in more advanced settings incorporating market incompleteness, jumps and transaction costs, among others.

Derivative valuation in the context of continuously trading markets was initiated by the seminal papers of \cite{black1973pricing} and \cite{merton1973theory}. The authors solved associated partial differential equations (PDEs) and obtained the price function of a European option in closed form, when the underlying asset follows a geometric Brownian motion (GBM). Their work (i.e. Black--Scholes--Merton formula) laid the foundation for the development of derivative pricing. Their results have been extended in many directions; most relevant to this study are extensions to the pricing of many other types of options, such as American options (see \cite{bjerksund1993closed}) lookback options (see \cite{goldman1979path}) and geometric average Asian options (see \cite{kemna1990pricing}), to mention a few. The distinctive exercise rights and structure of payoff reflect the complexity of financial derivatives.

Our paper focuses on the benefits of derivatives from the portfolio investment perspective for a maximized expected utility (EUT) agent. Investment incorporating derivatives have been studied from multiple perspectives. \cite{haugh2001asset} found a buy-and-hold strategy that minimizes the mean-squared distance to the terminal wealth of \cite{merton1969lifetime} continuously rebalancing portfolio. Moreover, an elasticity approach was introduced in \cite{kraft2003elasticity}, by which the author obtained the optimal strategy of a portfolio with path-dependent options. \cite{mi2020optimal} studied a portfolio of volatility derivatives (options or swaps) for a constant absolute risk aversion (CARA) investor. In additions, \cite{liu2003dynamic} investigated the optimal portfolio in a stock-derivatives market with Heston's SV model and jumps. Their results demonstrated the improvement in performance when using derivatives to complete the market, while showing that an infinite number of derivatives can be used with the same optimal performance for the portfolio. In contrast to the existing literature, our paper investigates how to best select derivatives. 



For this purpose, in addition to a maximized EUT, we develop a second optimization criterion for the portfolio manager as a way of selecting the best derivative. We choose risky asset exposure minimization motivated by two facts. First, the maximization of the utility can be achieved by an infinitely number of equally optimal derivatives (as per \cite{liu2003dynamic}). Second, minimizing risk exposure, as a second optimization criterion, aligns with practical needs in the industry. In general, investment companies face many constraints in the construction of their portfolios, many of which are imposed by regulatory agencies. The key factor behind regulatory constraints is the intention to control the exposure of a portfolio to risky assets, protecting an investor’s capital in the case of a market crash. Some of these risks are difficult to accurately model, which highlights the importance of minimizing exposure.  

Our findings demonstrate that derivatives can be used to reduce risk exposure with no impact on the level of satisfaction of the investor (e.g. maximum utility). We investigate the selection of derivatives in three specific option classes: (i) American, European and Asian calls and puts; (ii) American, European and Asian synthetic straddles; and (iii) basket options. We further compare one-asset and multi-asset options in various realistic situations, and we consider the relationship between risky asset exposure and portfolio rebalancing frequency. 

The contributions of the paper are summarized as follows:
\begin{enumerate}
    \item Given the infinite number of choices of equally optimal financial derivatives for an EUT investor, we explore an additional optimization criterion, namely, risk exposure minimization, to help investors make a practical derivative selection. 
    \item We demonstrate, in the context of two one-factor (e.g. GBM) assets, that the minimum number of derivatives needed not only to maximize EUT performance, but also to minimize risk exposure is exactly two. 
    \item In a comparison of the most popular types of simple one-asset options (e.g. American, European and Asian calls and puts), we illustrate that the deepest out-of-the-money Asian products are the best choices for minimizing risk exposure.   
    \item To avoid illiquid out-of-the-money options, which also require plenty of rebalancing, we explore optimal selections among straddles. We demonstrate the existence of an optimal strike price for risk exposure minimization, which is likely a better practical choice than out-of-the-money calls and puts.  
    \item Given the setting of two assets in the portfolio, we study the optimality of multi-asset derivatives. We determine that a basket option could be a better selection than one-asset Asian calls and puts in many realistic situations.  
    \item Several analyses are performed to solidify our findings; in particular, the relationship between risky asset exposure and portfolio rebalancing frequency is investigated, and the results are put to the test for a variety of parametric choices.
\end{enumerate}


The paper is organized as follows. Section 2 describes the EUT
problem for an investor allocating directly among financial derivatives. Given
the vast number of optimal derivatives available for this problem, Section 2
defines a criterion based on an additional optimization problem that aids in 
selecting a single optimal solution. Section 3 then explores two cases of
one-asset options available to a portfolio investor: (1) calls and puts and (2) straddles. Thereafter, Section 4 focuses on the benefit of multi-asset derivatives such as  basket options, and Section 5 concludes this study. Appendix \ref{chp5_appendix} presents all the proofs and complementary analyses in support of our findings.

\section{Mathematical setting and results}

Let $(\Omega ,\mathcal{F},\mathbb{P})$ be a complete probability space with
a right-continuous filtration $\{\mathcal{F}_{t}\}_{t\in \lbrack 0,T]}$. We
consider a frictionless market, in which trading occurs without transaction costs or market impact, comprising a money market account $M_{t}$ and two risky
assets $\bm{S_t}=[S_{t}^{(1)},S_{t}^{(2)}]^{T}$, with the following dynamics:
\begin{equation}
\begin{cases}
\frac{\mathrm{d}M_{t}}{M_{t}}=r\mathrm{d}t \\
\mathrm{d}\bm{S_{t}}=diag(\bm{S_{t}})\left[(r\cdot \bm{\mathbbm{1}}+diag(\bm{\sigma })\bm{\Lambda} )\mathrm{d}t+diag(\bm{\sigma}
)\mathrm{d}\bm{B_{t}}\right]%
\end{cases}
\label{chp5_bsm_stock}
\end{equation}%
where $\bm{B_{t}}=[B_{t}^{(1)},B_{t}^{(2)}]^{T}$ are Brownian motions modelling
the risk of two underlying assets, whose correlation is denoted by $\rho \in
\left( -1,1\right) $. Here, $r$ is the risk-free rate and $\bm{\mathbbm{1}}$ denote the
vector of ones, $\bm{\Lambda} =[\lambda ^{(1)},\lambda ^{(2)}]^{T}$ are constants
capturing the market price of risk of $\bm{B_{t}}$, and $\bm{\sigma} =[\sigma
^{(1)},\sigma ^{(2)}]^{T}$ represents the volatility of the two underlying
assets.

We now introduce a set of admissible financial derivatives on the assets $%
\bm{S_{t}}$, for a fixed $n\geq 1$:%
\begin{equation*}
\Omega _{O}^{(n)}=\left\{
\bm{O_{t}}=[O_{t}^{(1)},O_{t}^{(2)},...,O_{t}^{(n)}]^{T}\mid O_{t}^{(i)}\neq 0%
\text{, }i=1,...,n\text{ and rank}\left( \bm{\Sigma _{t}}\right) =2, \, t\in \lbrack 0,T] \right\}
\end{equation*}
where $\bm{\Sigma _{t}}$ represents the variance matrix of $O_{t}$. The
element $(i,j)$, $i=1,..,n$, $j=1,2$, of $\bm{\Sigma _{t}}$, denoted by $%
f_{t}^{ij}$,  represents the sensitivity of $O_{t}^{(i)}$ to the underlying
asset $S_{t}^{(j)}$, i.e. $f^{ij}=\frac{\partial O_{t}^{(i)}}{\partial S_{t}^{(j)}}\frac{1}{O_{t}^{(i)}}%
S_{t}^{(j)}\sigma ^{(j)}$. Note that $\Omega _{O}$ is an infinite set, which could contain standardized exchange-traded options and non-standardized
OTC options available to a generic investor. The reader
should observe that $f_{t}^{ij}$ depends on derivative type, style,
underlying price, strike price, time to maturity and other factors, and $%
\Sigma _{t}$ is a full-rank matrix, which allows us to continue working in
a complete market if $n\geq 2$. For simplicity, we also assume that the
derivatives in $\Omega _{O}^{(n)}$ will be rolled over, always keeping 
the same time to maturity and a non-zero value.

We are now ready to create a derivatives-based dynamic portfolio choice
problem for a risk-averse investor. The investor preference is measured by the  widely used and algebraically simple CRRA utility\footnote{The can  easily be extended to other utility functions. }.   We assume that an investor
allocates in an element of $\Omega _{O}$; that is, a specific $%
\bm{O_{t}}=[O_{t}^{(1)},O_{t}^{(2)},...,O_{t}^{(n)}]^{T}$ $(n\geq 2)$. Note that
by arbitrage arguments, the dynamics of the derivatives-based assets and the
market account are as follows:
\begin{equation}
\begin{cases}
\frac{\mathrm{d}M_{t}}{M_{t}}=r\mathrm{d}t \\
d\bm{O_{t}}=diag(\bm{O_{t}})\left[(r\cdot \bm{\mathbbm{1}}+\bm{\Sigma _{t}\Lambda} )\mathrm{d}t+\bm{\Sigma}
_{t}\mathrm{d}\bm{B_{t}}\right]%
\end{cases}
\label{chp_bsm_opt}
\end{equation}

The investor is not prohibited from trading on the underlying assets.
This would be equivalent to setting $n=2$ and taking $\bm{O_{t}}=\bm{S_{t}}$.

Let $\Omega _{\pi }^{(O)}$ denote the space of admissible strategies
satisfying the standard conditions, where the element $\bm{\pi _{t}}=[\pi
_{t}^{(1)},\pi _{t}^{(2)},...,\pi _{t}^{(n)}]^{T}$ represents the proportions
of the investor's wealth in the options $%
\bm{O_{t}}=[O_{t}^{(1)},O_{t}^{(2)},...,O_{t}^{(n)}]^{T}$ with the remaining $1-\mathbbm{1}%
^{T}\bm{\pi _{t}}$ invested in the cash account $M_{t}$. The wealth process $%
W_{t} $ satisfies
\begin{equation}
\frac{\mathrm{d}W_{t}}{W_{t}}=(r+\bm{\pi _{t}}^{T}\bm{\Sigma _{t}\Lambda} )\mathrm{d}t+\bm{\pi _{t}}^T\bm{\Sigma
_{t}}\mathrm{d}\bm{B_{t}}.
\end{equation}%
A CRRA utility function represents the investor's preference on the terminal
wealth $W_{T}$, which is given by
\begin{equation}
U(W_{T})=\frac{W_{T}^{1-\gamma }}{1-\gamma },
\end{equation}%
Moreover, $\gamma >0$, $\gamma \neq 1$ measures the investor's level of risk aversion. The investor's
objective is to derive an investment strategy $\bm{\pi _{t}^{\ast }}$ that
maximizes the EUT of the terminal wealth $W_{T}$. Then, the
investor's problem can be written as follows:
\begin{equation}
V(t,W)=\max_{\bm{\pi _{s\geq t}}\in \Omega _{\pi }^{(O)}}\mathbb{E}(U(W_{T})\mid
\mathcal{F}_{t}),
\end{equation}%
and $V(t,W)$ denotes the value function at time $t$. According to the
principles of stochastic control, we state the Hamilton-Jacobi-Bellman
 (HJB) equation for the value function $V$:
\begin{equation}
\sup_{\bm{\pi _{t}}}\left\{ V_{t}+W_{t}V_{W}(r+\bm{\pi _{t}^{T}\Sigma _{t}\Lambda} )+%
\frac{1}{2}W_{t}^{2}V_{WW}(\bm{\pi _{t}^{T}\Sigma _{t}\Phi \Phi ^{T}\Sigma
_{t}^{T}\pi _{t}})\right\} =0,  \label{chp5_hjb}
\end{equation}%
where $\bm{\Phi} =\left[
\begin{array}{cc}
1 & 0 \\
\rho & \sqrt{1-\rho ^{2}}%
\end{array}%
\right] $.

\bigskip

\begin{proposition}[Solution for $V$ and $\protect\bm{\pi _{t}^{*}}$]
The value function has the representation:
\begin{equation}
V(t,W)=\frac{W^{1-\gamma }}{1-\gamma }\exp \left( (1-\gamma )\left( r+\frac{1%
}{2}\frac{\bm{\Lambda} ^{T}(\bm{\Phi \Phi} ^{T})^{-1}\bm{\Lambda} }{\gamma }\right)
(T-t)\right) . \label{chp5_merton_valfun}
\end{equation}%
Moreover, $\bm{\pi _{t}^{\ast }}$ is an optimal strategy if it satisfies $\bm{\Sigma
_{t}}^{T}\bm{\pi _{t}^{\ast }}=\bm{\eta _{t}^{\ast }}$, where $\bm{\eta _{t}^{\ast }}$ is
given by
\begin{equation}
\bm{\eta _{t}^{\ast }}=\left[
\begin{array}{c}
\eta _{t}^{(1)} \\
\eta _{t}^{(2)}%
\end{array}%
\right] =\frac{(\bm{\Phi \Phi }^{T})^{-1}\bm{\Lambda }}{\gamma }.
\end{equation}%
\begin{proof}
See Appendix \ref{chp5_appendixa}.
\end{proof}\label{chp5_propos1}
\end{proposition}

\bigskip

Proposition \ref{chp5_propos1} highlights three important implications. First, for
any given element in $\Omega _{O}^{(n)}$, if $n>2$ numerous, indeed infinitely many,  strategies, all produce the same maximum value function. This can be interpreted as a redundant market case. Second, if $n=2$,  then a unique
optimal strategy exists for the problem. Finally, if $n=1$, there is no optimal solution: the value function cannot reach the global maximum; this is actually an incomplete market situation.\\

In summary, as there are a host of tradeable derivatives
in the financial market, hence a myriad of elements exist in $\Omega
_{O}^{(n)}$. This means there are an infinite number of choices of the portfolio
composition $\bm{O_{t}}$ that can deliver the same optimal solution to the EUT problem. The next section takes advantage of this pool of optimal
solutions to design a criterion that allows investors to select the best
portfolio composition, with the corresponding strategy. This extra criterion
is motivated by investor needs. Adding such a criterion will lead to an additional
optimization problem, the solution of which is explored below.

\subsection{Derivative selection criterion: minimizing $\ell_1$ risk exposure}

In this section, we propose a derivative selection criterion. Proposition %
\ref{chp5_propos1} illustrates that, given $\bm{O_{t}}$ $\in \Omega _{O}^{(n)}$,
with $n\geq 2$,  an optimal strategy $\bm{\pi _{t}^{\ast }}\in \Omega
_{\pi }^{(O)}$ exists that maximizes the EUT of terminal wealth. From the traditional dynamic portfolio choice perspective,  derivative
selection does not benefit the investor, because regardless of the
derivative chosen, the optimal strategy always achieves the same EUT.

In reality, investors are always concerned with the size of their
risky allocations. For example, market conditions may change over time, and a 
risky investor could suffer large, unexpected losses especially during crisis
periods. Regulatory constraints also force the investor to keep increasingly
large percentages of wealth in cash. This means that strategies with smaller
exposure on the risky products are naturally preferable to the investor. In
this regard, we design a simple derivative selection criterion aimed at
capturing this practical dilemma:
\begin{equation}
\underset{{\bm{O_{t}}\in \Omega _{O}^{(n)}}}{\min }\left\Vert \underset{\bm{\pi
_{s\geq t}}\in \Omega _{\pi }^{(O)}}{\arg \max }\mathbb{E}(U(W_{T})\mid
\mathcal{F}_{t})\right\Vert _{1}  \label{linear_pron}
\end{equation}%
where $\left\Vert \bm{\pi _{s\geq t}}\right\Vert
_{1}=\sum\limits_{i=1}^{n}\left\vert \pi _{t}^{(i)}\right\vert $ represents
the $\ell _{1}$ norm of allocations at time $t$. Note that this objective is equivalent to maximizing the cash position while shorting less. 


As we mentioned before, it is suboptimal for the investor to choose a
portfolio composition size $n<2$ based on the unhedgeable
risk resources and incompleteness of the market. However, the investor might
be interested in a redundant $(n>2)$ market situation,  hoping to reduce their risky asset exposure.  In the next proposition, we demonstrate that the best choice of $n$, for problem \eqref{linear_pron}, is actually $n=2$.

\bigskip

\begin{proposition}
Assume that an optimal solution for problem \eqref{linear_pron} exists for $n\geq 2
$, then, \eqref{linear_pron} leads to the same minimal $\ell _{1}$ norm for any $n\geq 2$. In addition, an optimal strategy exists for problem 
\eqref{linear_pron} such that the number of non-zero allocations is less than or equal to $2$.

\begin{proof}
See Appendix \ref{appendix_twoproblem}.

\end{proof}\label{chp5_propos_twoproblem}
\end{proposition}

Proposition \ref{chp5_propos_twoproblem} demonstrates that redundancy will not offer any additional help with the investor's risky asset exposure. In other words,
working with $n=2$ is sufficient for problem \eqref{linear_pron}. This
allows us to work with the simplest case given a complete market setting (i.e. $n=2$).

\section{Applications to one-asset derivatives}

In this section, we solve the derivative selection problem---that is, %
\eqref{linear_pron}---for $n=2$, within subsets of the derivative set ${\Omega
_{O}^{(2)}}$.  The derivative selection problem is rewritten as
\begin{equation}
\underset{{\bm{O_{t}}\in \Omega _{O}^{(2,1)}}}{\min }\left\Vert \underset{\bm{\pi
_{s\geq t}}\in \Omega _{\pi }^{(O)}}{\arg \max }\mathbb{E}(U(W_{T})\mid
\mathcal{F}_{t})\right\Vert _{1},
 \label{linear_pro_21}
\end{equation}
where  ${\Omega _{O}^{(2,1)}}$ captures one-asset (single-stock) derivatives, which can be represented as follows:
\begin{equation*}
\Omega _{O}^{(2,1)}=\left\{\bm{ O_{t}}=[O_{t}^{(1)},O_{t}^{(2)}]^{T}\mid O_{t}\in
\Omega _{O}^{(2)}\text{, }O_{t}^{(i)}=g\left( S^{(i)}\right) \text{, }%
i=1,2\text{, } t\in \lbrack 0,T] \right\}.
\end{equation*}
In this situation, the option's variance matrix for a portfolio composition $\bm{O_t}=[O_{t}^{(1)},O_{t}^{(2)}]^{T}\in\Omega _{O}^{(2,1)}$ is defined by
\begin{equation}
\bm{\Sigma_t}=\left[\begin{array}{cc}
f^{(1)}_t& 0 \\
0 & f^{(2)}_t
\end{array}\right],
\label{chp5_variance_matrix_single}
\end{equation}
which, provided $f^{(1)}_t$ and $f^{(2)}_t$ are nonzero, is a rank $2$ (non-singular) diagonal matrix.  By It\^{o}'s lemma, the sensitivity of  $O_t^{(i)}$ is a function of the option Delta, spot price $S_t^{(i)}$, stock volatility $\sigma^{(i)}$ and the option price $O_t^{(i)}$:
\begin{equation}
 f^{(i)}  = \frac{\partial O_t^{(i)}}{\partial S_t^{(i)}}\frac{S_t^{(i)}\sigma^{(i)}}{O_t^{(i)}}.
 \label{chp5_option_vol}
\end{equation}
The next proposition states the fundamental principle of one-asset option selection.

\begin{proposition}[Fundamental principle of one-asset option selection]  A portfolio composition:\\ $\bm{O^*_t}=[O_t^{(1),*},O_t^{(2),*}]^T\in\Omega _{O}^{(2,1)}$ is optimal for problem \eqref{linear_pro_21} if and only if
\begin{equation}
    O_t^{(i),*}=\underset{O_t^{(i)}}{\arg \max }\left\vert \frac{\partial O_t^{(i)}}{\partial S_t^{(i)}}\frac{1}{O_t^{(i)}}\right\vert.
\end{equation}
\begin{proof}
See Appendix \ref{appendix_single}.
\end{proof}\label{chp5_propos_single}
\end{proposition}
Note that both option Delta and price must be bounded away from $0$ and $\infty$ to avoid the suboptimal incomplete market case, such that the option sensitivity $f^{(i)}\in(0,\infty)$ and variance matrix $\Sigma_t$ is non-singular. In other words, we could witness an infinitely large allocation as $f^{(i)}\rightarrow0$ (to be explained in Section \ref{chp5_4.1}), which would also lead to a suboptimal solution (incomplete market) and hence a departure from the investor's target (i.e. maximizing utility and minimizing risk exposure).

The fundamental principle of one-asset option selection illustrates that the selection of one-asset options is separable. Investors can first pick the option with the largest relative sensitivity to $S_t^{(1)}$ among all one-asset options on $S_t^{(1)}$ as $O_t^{(1),*}$, and they can then select $O_t^{(2),*}$ in a similar way. Based on this principle, we consider the case where $\Omega _{O}^{(2,1)}$ is a subset of put and call options in Section \ref{chp5_4.1}. Then, the best option style and strike price for minimizing $\left\Vert\bm{\bm{\pi_t}}\right\Vert_1$ is quantified.

Selection of put and call options is studied first due to their popularity. Nonetheless, calls and puts have a problem: their optimal is on the boundary of the strike price range. This could lead to illiquid choices (high out-of-the-money options) or, even worse, incomplete market suboptimality on the limit as the strike price goes to zero (puts) or infinity (calls). In Section \ref{chp5_4.2}, we investigate derivative selection in a subset of straddles. Straddles are also popular products, which avoid the boundary optimality of calls and puts. Note that the selection of $O_t^{(1),*}$ and $O_t^{(2),*}$ is independent, and the procedures are similar; hence, for simplicity, we only present the result for $O_t^{(2),*}$. 

\bigskip

The chosen parameters are presented in Table \ref{para_merton_chp5} \footnote{See Appendix \ref{chp5_compare_one_multi_opt} for analysis of other parameter choices.}. These parameters are considered to be plausible.

 \begin{table}[ht]\centering\caption{Parameter Value}\label{para_merton_chp5}
\begin{tabular}{llll}
\toprule
Parameter &Value& Parameter &Value \\\midrule
$\sigma^{(1)}$& 0.13 &$\sigma^{(2)}$ &0.2 \\
$r$ & 0.05 & $\rho$ & 0.4 \\
$\lambda^{(1)}$ & 0.52& $\lambda^{(2)}$ & 0.6 \\
Investment horizon $T$ &1& Time to maturity of options $\hat{T}$ & 2
\\ $S^{(1)}_0$ &40& $S^{(2)}_0$ &30\\
$\eta_t^{(1)}$ &0.083 &$\eta_t^{(2)}$ &  0.117 \\
$\gamma$ & 4.0& &\\
\botrule
\end{tabular}
\end{table}

The option variance matrix may not be solvable in closed form. Therefore, we approximate the sensitivity of European, American and arithmetic average Asian options via the finite difference method: the Delta of option $O_t$ is given by
\begin{equation}
Delta= \frac{O_t(S+\Delta S)-O_t(S-\Delta S)}{2\Delta S}.
\end{equation}
Here, $O_t(S)$ is the simulated option price given spot price $S$. In addition, we estimate the sensitivity of American options with the generalized infinitesimal perturbation analysis approach introduced by \cite{chen2014american}.

\subsection{Put and call options}
\label{chp5_4.1}
We first consider the derivative selection problem \eqref{linear_pro_21} on the subset denoted by $\Omega _{O}^{(2,put\ call)}$ that contains only European-, American- and Asian-style put and call options:
\begin{equation*}
\begin{split}
\Omega _{O}^{(2,put\ call)}=\left\{ \bm{O_{t}}=[O_{t}^{(1,j)},O_{t}^{(2,j)}]^{T}\mid O_{t}^{(i,j)}=g^{(j)}\left( S^{(i)}\right) \text{, }i=1,2, \, t\in \lbrack 0,T]
 \right\}.   
\end{split}
\end{equation*}
For both practical and theoretical reasons, the strike price $K^{(i,j)}$ of a given option $O_t^{(i,j)}$ is bounded within $[A^{(i,j)},B^{(i,j)}]$, where $j\in\{\textit{Euro Call, Asian Call, Amer Call, Euro Put, Asian Put, Amer Put}\}$. The put option strike price is bounded away from $0$; that is, $A^{(i,j)}>0$, where $j\in\{\textit{Euro Put, Asian Put, Amer Put}\}$. Similarly, the call option strike excludes $\infty$. Both conditions ensure the non-zero option price assumption. For simplicity, we also assume that all the options have the same time to maturity $\hat{T}=2$, and we search the optimal portfolio composition in terms of option type, style and strike price.

European put and call prices and their sensitivities are solved by the well-known Black–Scholes–Merton model (see \cite{black1973pricing}). Let $O_t^{(i,Euro\ Call)}$ and $O_t^{(i,Euro\ Put)}$ be a call and put option on $S_t^{(i)}$ given in \eqref{chp5_bsm_stock}. The Black–Scholes–Merton model indicates that
\begin{equation}
\begin{split}
O_t^{(i,Euro\ Call)}&=S_t^{(i)}N(d_1)-K^{(i,Euro\ Call)}e^{-r(\hat{T}-t)}N(d_2)\\
O_t^{(i,Euro\ Put)}&=K^{(i,Euro\ Put)}e^{-r(\hat{T}-t)}N(-d_2)-S_t^{(i)}N(-d_1)\\
\frac{\partial O_t^{(i,Euro\ Call)}}{\partial S_t^{(i)}}&=N(d_1) \quad\quad\quad \frac{\partial O_t^{(i,Euro\ Put)}}{\partial S_t^{(i)}}=-N(-d_1) 
\end{split}\label{chp5_euro_price}
\end{equation}
where $N$ is the cumulative distribution function of a standard normal random variable and 
\begin{equation}
d_1=\frac{\ln{(S_t^{(i)}/K^{(i,Euro\ Call/Put)})}+(r+\frac{1}{2}(\sigma^{(i)})^2)(\hat{T}-t)}{\sigma^{(i)}\sqrt{\hat{T}-t}}, \quad d_2=d_1-\sigma^{(i)}\sqrt{\hat{T}-t}.
\label{ch5_bsm_d}
\end{equation}

The proposition next demonstrates the existence of an optimal portfolio composition $O_t^{\ast}\in\Omega _{O}^{(2,put\ call)}$, given the assumption that $O_t^{(i,j)}\in\mathbb{C}^1$ and the Delta is non-zero. 

\begin{proposition}[Existence of optimal portfolio composition in put and call subset]
 Assume $O_t^{(i,j)}\left(S^{(i)},K^{(i,j)}\right)\in \mathbb{C}^1$ on $(0,\infty)\times [A^{(i,j)},B^{(i,j)}]$\footnote{$\frac{\partial O_t^{(i,j)}}{\partial K^{(i,j)}}$ on boundary is defined as the one-sided derivative.}, the optimal portfolio composition for problem \eqref{linear_pro_21} within the subset $\Omega _{O}^{(2,put\ call)}$ exists.

\begin{proof}
See Appendix \ref{appendix_putcall_exist}.
\end{proof}\label{chp5_propos_putcall_exist}
\end{proposition}

It is easy to show the expression for this optimal composition, in the case of the European options, as clarified in the following corollary. 

\begin{corollary}
The risk exposure to an European call option decreases with $K^{(i,Euro\ Call)}$ and converges to $0$ as $K^{(i,Euro\ Call)}\rightarrow \infty$. Similarly, the risk exposure to an European put option increases with $K^{(i,Euro\ Call)}$ and converges to $0$ as $K^{(i,Euro\ Put)}\rightarrow 0$.\\ 
Therefore, the optimal European call is achieved when $K^{(i,Euro\ Call)}=B^{(i,Euro\ Call)}$ and the optimal European put is achieved when $K^{(i,Euro\ Put)}=A^{(i,Euro\ Put)}$ .

\begin{proof}
See Appendix \ref{appendix_limit}.
\end{proof}\label{chp5_propos_limit}
\end{corollary}

Corollary \ref{chp5_propos_limit} demonstrates that $\left\Vert\bm{\bm{\pi_t}}\right\Vert_1$ can vanish if the European call option's strike price approaches infinity. These extreme options cannot be found in the market, but more importantly, they would also lead to a violation of the non-singular matrix $\Sigma_t$ condition, creating an incomplete market situation. This is known as the problem of ``boundary optimality.'' The corollary thus illustrates the importance of our derivative selection. 

\begin{figure}[ht]
	\centering
	\vspace{0.35cm}
	\subfigtopskip=2pt
	\subfigbottomskip=2pt
	\subfigcapskip=-5pt
	\subfigure[allocation on call options]{
		\includegraphics[width=0.45\linewidth]{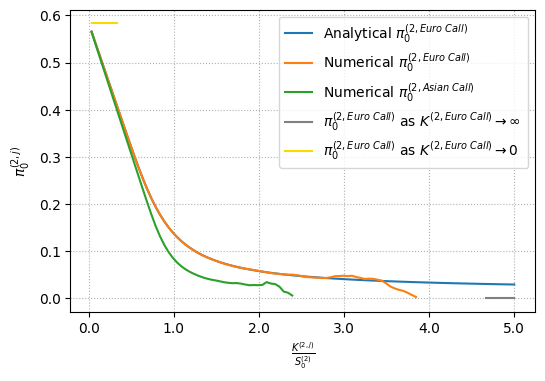}}
	\quad
	\subfigure[allocation on put options]{
		\includegraphics[width=0.45\linewidth]{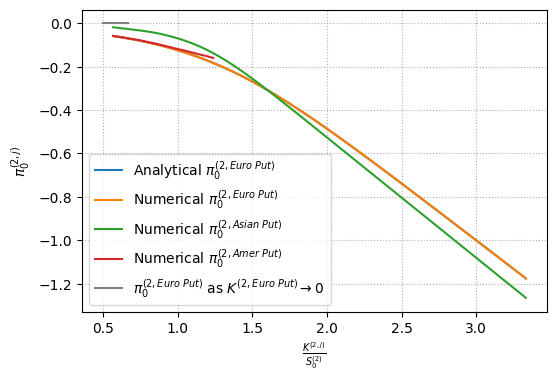}}
		\subfigure[Derivative selection region]{
		\includegraphics[width=0.45\linewidth]{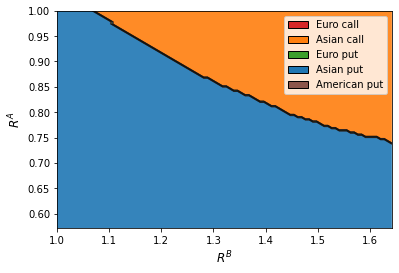}}

	\caption{Allocation on options versus strike price}
    \label{chp5_pi_vs_K}
\end{figure}

Figure \ref{chp5_pi_vs_K} (a) exhibits $\pi^{(2,j)}_0$---that is, the allocation on call options $O_t^{(2,j)}$ at $t=0$---as a function of strike price and spot price ratio $K^{(2,j)}/S_t^{(2)}$; for example, $K^{(2,j)}/S_t^{(2)}=1$ indicates at-the-money options. The analytical sensitivity of European calls is obtained with \eqref{chp5_option_vol} and \eqref{chp5_euro_price}, with which the optimal strategy is known immediately. Note that the European call option reduces to $S_t^{(2)}$ when $K^{(2,Euro\ Call)}=0$, and the allocation hence converges to the optimal strategy on $S_t^{(2)}$. The allocation on the European call obtained by the finite difference method is also plotted to verify the accuracy of numerical approximation.  Allocations obtained via analytical and numerical approaches visually overlap except when $K^{(2,Euro\ Call)}$ is extremely large. The allocation to Asian calls is similar to that of European calls when the strike prices are small, but it decreases faster. Asian calls are consequently preferable to European calls given the same upper bound on strike price; that is, $B^{(2,Euro\ call)}=B^{(2,Asian\ call)}$.

The allocation on put options is illustrated in Figure \ref{chp5_pi_vs_K} (b). The absolute allocations for European, American\footnote{We assume stocks which pay no dividends, so American call is identical to the European call in (a). In addition, the allocation on American put  is shown in (b) when $K^{(2,Amer\ Put)}<37.7$, i.e. the region, in which the American put should be held and not yet exercised.} and Asian puts increase with strike price because of the decreasing instantaneous sensitivity. The absolute allocation on Asian puts is the smallest when $K^{(2,Asian\ Put)}$ is small, but the opposite occurs as the strike price rises. Figure \ref{chp5_pi_vs_K} not only confirms Corollary \ref{chp5_propos_limit}---that is $\pi_t^{(2,Euro\ Call)}\rightarrow0$ as $K^{(2,Euro\ Call)}\rightarrow\infty$ and $\pi_t^{(2,Euro\ Put)}\rightarrow0$ as $K^{(2,Euro\ Put)}\rightarrow0$---but also demonstrates a similar conclusion for Asian and American options.

In summary, the absolute allocation on a put or call option is monotone with strike price; hence, the optimal choice of $O_t^{(2,j)}$ is on the boundary. This means we only need to compare the allocation $\pi^{(2,j)}$ on the deepest out-of-the-money option to obtain the option with the smallest absolute allocation.

Next, we consider the case where options of different style\footnote{The actual range of American put strike price is the intersection of  $[A^{(i,Amer\ Put)},B^{(i,Amer\ Put)}]$ and the region of $K^{(i,Amer\ Put)}$ such that the option is not exercised immediately, hence American option is not considered when $[A^{(i,Amer\ Put)},B^{(i,Amer\ Put)}]$ is mutually exclusive with that region. } share an identical boundary:
\begin{equation*}
\begin{split}
  &[S_t^{(2)},R^BS_t^{(2)}]=[A^{(2,j)},B^{(2,j)}], \quad j\in\{Euro\ Call,Asian\ Call\}\\
&[R^AS_t^{(2)},S_t^{(2)}]=[A^{(2,j)},B^{(2,j)}], \quad j\in\{Euro\ Put,Asian\ Put,Amer\ Put\}.
\end{split}
\end{equation*}
 Here, $R^{A}\leq1$, and $R^{B}\geq1$. As we move the lower bound of put option strike price $R^AS_t^{(2)}$ and the upper bound of call option strike price $R^BS_t^{(2)}$, the optimal choice of $O_t^{(2,j)}$ is shown in Figure \ref{chp5_pi_vs_K} (c). The Asian option is preferable compared to American and European options. Asian calls dominate when both $R^{A}$ and $R^B$ are large, whereas Asian puts dominate when both $R^{A}$ and $R^B$ are small. 

\begin{figure}[ht]
	\centering
	\vspace{0.35cm}
	\subfigtopskip=2pt
	\subfigbottomskip=2pt
	\subfigcapskip=-5pt
	
    \includegraphics[width=0.6\linewidth]{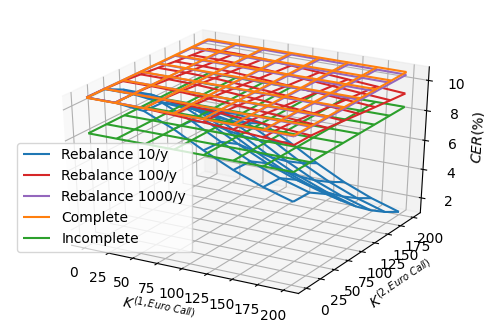}
	
	\caption{Certainty equivalent rate (CER) versus strike price with different rebalancing frequencies}
    \label{GBM_vafun_callvscall_K}
\end{figure}

Figure \ref{GBM_vafun_callvscall_K} displays the performance of the portfolio $\bm{O_t}=[O_t^{(1, Euro\ Call)},O_t^{(2, Euro\ Call)}]^T$ versus strike prices for different rebalancing frequencies. The portfolio performance is measured by annualized certainty equivalent  (CER), defined as
\begin{equation}
U(W_0\exp{(CER*T)})=V(0,W_0).
\end{equation} 
The theoretical optimal CER (orange wireframe) is plotted as the benchmark, which can only be achieved by continuously rebalancing. We also present the incomplete market CER (i.e. the green wireframe) for comparison purposes, for example obtained through the lack of an asset to hedge the risk in $B_t^{(2)}$. A portfolio with in-the-money call options is insensitive to the rebalancing frequency, and the loss from occasional rebalancing is not significant. On the other hand, the CER of a portfolio with two deep out-of-the-money calls could be even smaller than that of an incomplete market CER when the portfolio is only rebalanced $10$ times per year, whereas it approaches complete market CER as the rebalancing frequency increases.\\\ 
This is important for investors reducing their risk exposure with deep out-of-the-money options. These products lack liquidity; hence, the trading strategy might fail because investors cannot adjust their position quickly enough. In summary, the best (out-of-the-money) options are those requiring more frequent rebalancing, as Figure \ref{GBM_vafun_callvscall_K} demonstrates. This points at future research addressing the trade-off between exposure and rebalancing. 

\subsection{Straddles}
\label{chp5_4.2}
Next, we consider the derivative selection in a subset of options called straddles:
\begin{equation*}
\Omega _{O}^{(2,straddle)}=\left\{ \bm{O_{t}}=[O_{t}^{(1,j)},O_{t}^{(2,j)}]^{T}\mid  O_{t}^{(i,j)}=g^{(j)}\left( S^{(i)}\right)\text{, } i=1,2  \text{, }  \, t\in \lbrack 0,T] \right\}.
\end{equation*}
where $j\in\{Euro\ Strad, Asian\ Strad, Amer\ Strad\}$. A straddle is an option synthesized by simultaneously taking a long position in a call and a put option; hence, the terminal payoff of a European straddle is $\left\vert S_T^{(i)}-K^{(i,Euro\ Strad)}\right\vert$, where $K^{(i,Euro\ Strad)}$ denotes the strike price. The European straddle price and Delta are obtained with the Black–Scholes– Merton model:
\begin{equation}
\begin{split}
O_t^{(i, Euro\ Strad)}&=S_t^{(i)}(2N(d_1)-1)-K^{(i,Euro\ Strad)}e^{-r(\hat{T}-t)}(2N(d_2)-1)\\
&\frac{\partial O_t^{(i,Euro\ Strad)}}{\partial S_t^{(i)}}=2N(d_1)-1.
\end{split}
\end{equation}
$d_1$and $d_2$ are defined in \eqref{ch5_bsm_d}. We substitute the straddle's Delta   into \eqref{chp5_option_vol}, and the sensitivity of straddle $O_t^{(i,Euro\ Strad)}$ is given by
\begin{equation}
 f^{(i,Euro\ Strad)}  = \frac{(2N(d_1)-1)S_t^{(i)}\sigma^{(i)}}{O_t^{(i,Euro\ Strad)}}.
 \label{chp5_euro_strad_sen}
\end{equation}
The non-singular variance matrix condition requires non-zero sensitivity $f^{(i,Euro\ Strad)}$, such that $2N(d_1)-1\neq0$ and the feasible region for the strike price is as follows:
\begin{equation}
\begin{split}
 K^{(i,Euro\ Strad)}&\in\left[0,S_t^{(i)}\exp{(r+\frac{1}{2}(\sigma^{(i)})^2)(\hat{T}-t)}\right )\\ &\cup\left(S_t^{(i)}\exp{(r+\frac{1}{2}(\sigma^{(i)})^2)(\hat{T}-t)},\infty\right).    
\end{split}
\end{equation}
It is easy to verify that $O_t^{( Euro\ Strad)}\in(0,\infty)$ and $O_t^{(i,j)}\left(S^{(i)},K^{(i,j)}\right)\in \mathbb{C}^1$.  We define the feasible region for the straddle $O_t^{(i,j)}$ by analogy:  $K^{(i,j)}\in[0,A^{(i,j)})\cup(A^{(i,j)},\infty)$, where $j\in\{Euro\ Strad, Asian\ Strad\}$, and $K^{(i,j)}\in[0,A^{(i,j)})\cup(A^{(i,j)},B^{(i,j)}]$, where $j\in\{Amer\ Strad\}$. Here, $A^{(i,Amer\ Strad)}$ is the point Delta of $O_t^{(i,Amer\ Strad)}$ equal to $0$, and $B^{(i,Amer\ Strad)}$ is the maximum strike price such that the American put is not immediately exercised. The next theorem shows the existence of an optimal straddle (i.e. an optimal strike price).

\bigskip

\begin{proposition}[Existence of optimal portfolio composition in the straddle subset]
There exists a portfolio composition $O_{t}^{\ast}=[O_{t}^{(1),\ast},O_{t}^{(2),\ast}]^{T}$ with finite strike prices $K^{(i,j,\ast)}$, such that  $O_{t}^{\ast}$ is optimal  for problem \eqref{linear_pro_21} within $\Omega _{O}^{(2,straddle)}$. Here the optimal strike price, $K^{(i,j,\ast)}$ is  the solution of the equation:
\begin{equation}
    \frac{\partial^2 O_{t}^{(i,j)}}{\partial S_t^{(i)} \partial K^{(i,j)}}O_{t}^{(i,j)}-\frac{\partial O_{t}^{(i,j)}}{\partial S_t^{(i)}}\frac{\partial O_{t}^{(i,j)}}{\partial K^{(i,j)}}=0.
\end{equation}
\begin{proof}
See Appendix \ref{appendix_straddle}.
\end{proof}\label{chp5_propos_straddle}
\end{proposition}

\begin{figure}[ht]
	\centering
	\vspace{0.35cm}
	\subfigtopskip=2pt
	\subfigbottomskip=2pt
	\subfigcapskip=-5pt
		\includegraphics[width=0.7\linewidth]{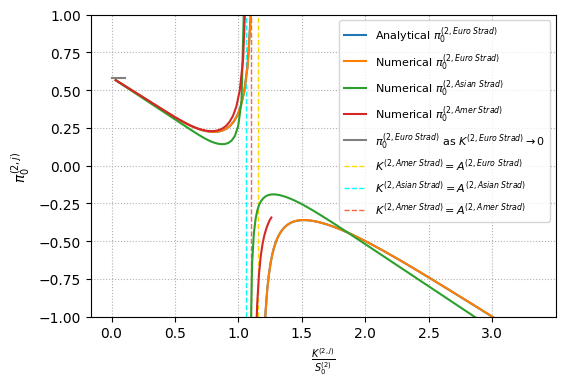}

	\caption{Allocation on straddle versus $\frac{K^{(i,j)}}{S_0^{(i,j)}}$}
    \label{GBM_straddle_pi}
\end{figure}

Next, we quantify the derivative selection within the subset of straddles. Here, we only illustrate the optimal choice of $O_t^{(2,j)}$ because of the separable selection within a one-asset option subset (see  Proposition \ref{chp5_propos_single}). We first compute the optimal strike price of American straddle $B^{(i,Amer\ Strad)}=37.7$ and the unfeasible point $A^{(2,j)}$ for European, Asian and American straddles; this is where the Delta of $O_t^{(2,j)}$ is equal to $0$. From the formulas above, we have $A^{(2,Euro\ Strad)}=S_t^{(2)}\exp{(r+\frac{1}{2}(\sigma^{(2)})^2)(\hat{T}-t)}=34.5$, $A^{(2,Asian\ Strad)}=31.9$ and $A^{(2,Amer\ Strad)}=32.9$, which are all obtained with Brent’s algorithm.

Figure \ref{GBM_straddle_pi} depicts the allocation $\pi_0^{(2,j)}$, where $j\in\{Euro\ Strad, Asian\ Strad, Amer\ Strad\}$ versus the ratio of spot price to strike price. The allocation $\pi_0^{(2,j)}$ has the shape of a hyperbola, and it approaches $\pm\infty$ as $K^{(2,j)}\uparrow\downarrow A^{(2,j)}$, which forms a ``cliff.'' In contrast to the put and call option, the optimal straddle is found in the interior regardless of the option style. The boundary optimality issue is thus  avoided, and it's plausible that the straddle minimum risk exposure will have acceptable liquidity. Moreover, the optimal straddle minimizing the risk exposure lies on the left branch; that is, $K^{(2,j)}\in[0,A^{(2,j)})$. The Asian straddle is superior to European and American straddles because of its smaller absolute allocation. 

The allocation on the European straddle $\pi_t^{(i,Euro\ Strad)}$ is solved in closed form with \eqref{chp5_euro_strad_sen}:
\begin{equation}
\begin{split}
     \pi_t^{(i,Euro\ Strad)}&=\frac{\eta_t^{(i)}}{f^{(i,Euro\ Strad)}}\\ &=\frac{\eta_t^{(i)}(2S_t^{(i)}(2N(d_1)-1)-K^{(i,Euro\ Strad)}e^{-r(\hat{T}-t)}(2N(d_2)-1)}{(2N(d_1)-1)S_t^{(i)}\sigma^{(i)}}, 
\end{split}
\end{equation}

which is a function of the ratio of strike price to spot price $K^{(i,Euro\ Strad)}/S_t^{(i)}$. 

\begin{figure}[ht]
	\centering
	\vspace{0.35cm}
	\subfigtopskip=2pt
	\subfigbottomskip=2pt
	\subfigcapskip=-5pt
	\subfigure[Optimal ratio $\frac{K^{(i,Euro\ Strad)}}{S_t^{(i,Euro\ Strad)}}$ versus time $t$]{
		\includegraphics[width=0.45\linewidth]{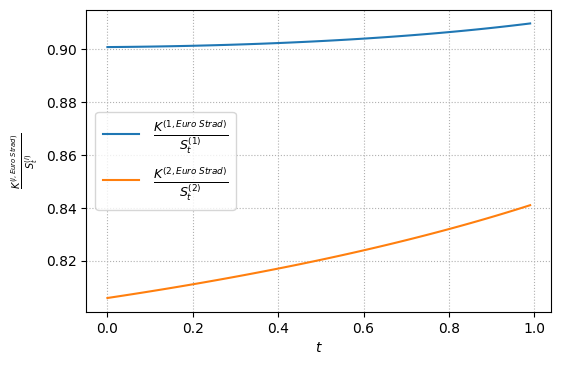}}
	\quad
	\subfigure[CER versus rebalancing frequency ]{
		\includegraphics[width=0.45\linewidth]{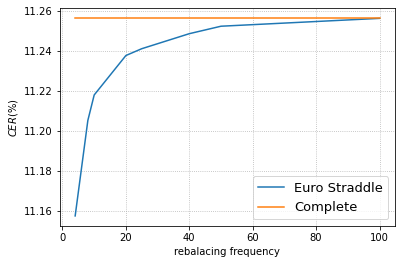}}
	\caption{Straddle analysis}
    \label{chp5_CER_straddle}
\end{figure}

In the subset of put and call $\Omega_O^{(2,put\ call)}$ (previous section), the optimal option is always found at the boundary. Therefore, investors stick to the option by rolling over with the same strike price. The optimal option in $\Omega_O^{(2,straddle)}$ depends on the spot price and time-dependent optimal ratio $K^{(i,Euro\ Strad)}/S_t^{(i)}$, and the investor should roll from the current holding to new straddles at each rebalancing time to minimize risk exposure. In this regard, Figure \ref{chp5_CER_straddle} (a) plots the optimal strike and spot price ratio of European straddle $K^{(i,Euro\ Strad)}/S_t^{(i)}$ versus time $t$. Both  $K^{(1,Euro\ Strad)}/S_t^{(1)}$ and $K^{(2,Euro\ Strad)}/S_t^{(2)}$ increase with time $t$, while  $K^{(2,Euro\ Strad)}/S_t^{(2)}$ grows faster. 

The connection between portfolio CER and rebalancing frequency is demonstrated in Figure \ref{chp5_CER_straddle} (b). As expected, the portfolio CER approaches the theoretical result as rebalancing frequency rises. Note that the CER of the rolling straddle portfolio is close to the theoretical CER even when the rebalancing frequency is less than 10 times per year, suggesting that rebalancing even relatively infrequently causes only a small loss. This is another benefit of choosing straddles over out-of-the-money calls or puts.

In conclusion, straddles are an ideal option class for two reasons. First, the optimal straddle for minimizing risk exposure happens to be an active and liquid option. In addition, a rolling straddle portfolio is insensitive to the rebalancing frequency, thus reducing an investor's additional costs, such as transaction costs (although transaction cost is not exactly modelled yet).

\section{Multi-asset option selection}
\label{chp5_5}

Multi-asset options are commonly traded in the OTC market. In this section, we explore the benefits of including such options in an investor's portfolio. A subset of multi-asset options is defined as follows:
\begin{equation*}
\begin{split}
\Omega _{O}^{(2,multi\ asset)}=\Big\{  \bm{O_{t}}&=[O_{t}^{(1,j_1)},O_{t}^{(2,j_2)}]^{T}\mid O_{t}^{(1,j_1)}=g^{(j_1)}\left( S^{(1)}\right) \text{, }   j_1= \textit{one-asset option;   }   \\    &O_{t}^{(2,j_2)}=g^{(j_2)}\left( S^{(1)},S^{(2)}\right) \text{, }   j_2= \textit{multi-asset option}, \, t\in \lbrack 0,T] \Big\}.
\end{split}
\end{equation*}
Assume the portfolio composition $\bm{O_{t}}\in\Omega _{O}^{(2,multi\ asset)}$ consists of a one-asset option and a multi-asset option. The variance matrix then has the representation
\begin{equation}
\Sigma_t=\left[\begin{array}{cc}
f^{(11)}_t& 0 \\
f^{(21)}_t & f^{(22)}_t
\end{array}\right].
\label{chp5_variance_matrix_multi}
\end{equation}
The next proposition states the fundamental principle of derivative selection in the subset of multi-asset options.

\begin{proposition}[Fundamental principle of multi-asset option selection]  If a portfolio composition $\bm{O^*_t}=[O_t^{(1),*},O_t^{(2),*}]^T$  is optimal for problem \eqref{linear_pro_21} within $\Omega _{O}^{(2,multi\ asset)}$, then
\begin{equation}
    O_t^{(1),*}=\underset{O_t^{(1,j_1)}}{\arg \max }\left\vert \frac{\partial O_t^{(1,j_1)}}{\partial S_t^{(1)}}\frac{1}{O_t^{(1,j_1)}}\right\vert.
\end{equation}
\begin{proof}
See Appendix \ref{appendix_multi}.
\end{proof}\label{chp5_propos_multi}
\end{proposition} 

Proposition \ref{chp5_propos_multi} demonstrates  a necessary condition for multi-asset option selection and reveals the sequential selection property for problem \eqref{linear_pro_21} within $\Omega _{O}^{(2,multi\ asset)}$. Investors should pick the one-asset option with the largest relative sensitivity to $S_t^{(1)}$ first, and they should then search for the optimal multi-asset option (see Equation \eqref{chp5_l1nrom_multi}) given a fixed $f^{(11)}_t$.

Now, we illustrate the multi-asset portfolio selection with an example of basket options. The subset of basket option is given by

\begin{equation*}
\begin{split}
\Omega _{O}^{(2,call\ basket)}&=\Big\{  \bm{O_{t}}=[O_{t}^{(1,j_1)},O_{t}^{(2,j_2)}]^{T}\mid O_{t}^{(1,j)}=g^{(j_1)}\left( S^{(1)}\right) \text{, }   j_1= \textit{European call;   }   \\    &O_{t}^{(2,j_2)}=g^{(j_2)}\left( S^{(1)},S^{(2)}\right) \text{, }   j_2= \textit{Basket Call or Basket Put}, \, t\in \lbrack 0,T] \Big\}.
\end{split}
\end{equation*}
Notably, $\Omega _{O}^{(2,call\ basket)}\subset\Omega _{O}^{(2,multi\ asset)}$. The basket option, simultaneously hedging the risk on a combination of two assets, has the following payoff\footnote{Investor can choose the weight on each asset of basket option in OTC market, we only consider equal weighted case in this paper. }:
\begin{equation}
\begin{cases}
O_T^{(2,j_2)}=\left(S_T^{(1)}+S_T^{(2)}-K^{(2,j_2)}\right)^+ & j_2= \textit{Basket Call}\\
O_T^{(2,j_2)}=\left(K^{(2,j_2)}-S_T^{(1)}-S_T^{(2)}\right)^+  & j_2= \textit{Basket Put}.
\end{cases}
\end{equation}
Furthermore, $K^{(i,j_i)}\in[A^{(i,j_i)},B^{(i,j_i)}]$, where $j_1=Euro\ Call$ and $j_2\in \{ Basket\ Call, Basket\ Put\}$, denotes the strike price of call and basket options. The existence of optimal portfolio composition within the subset of basket option is demonstrated in the next proposition.

\begin{proposition}[Existence of optimal portfolio composition in the subset of basket option]
Let the basket option price $O_t^{(2,j_2)}$ be a function of $\left(S^{(1)}_t,S^{(2)}_t,K^{(2,j_2)}\right)$ and $O_t^{(2,j_2)}\in \mathbb{C}^1$ on $(0,\infty)\times(0,\infty)\times [A^{(2,j_2)},B^{(2,j_2)}]$, then the optimal portfolio composition for problem \eqref{linear_pro_21} within the subset $\Omega _{O}^{(2,call\ basket)}$ exists.\\

\begin{proof}
See Appendix \ref{appendix_callbasket_exist}.
\end{proof}\label{chp5_propos_callbasket_exist}
\end{proposition}

\begin{figure}[ht]

	\centering
	\vspace{0.35cm}
	\subfigtopskip=2pt
	\subfigbottomskip=2pt
	\subfigcapskip=-5pt
	\subfigure[$\ell_1$ norm of allocation versus strike price]{
		\includegraphics[width=0.45\linewidth]{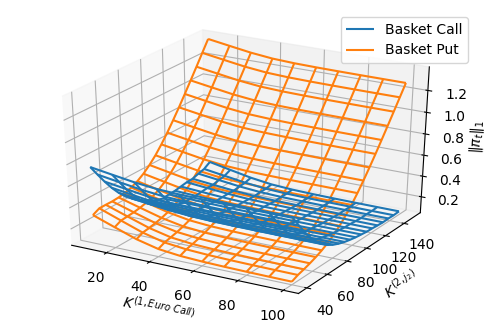}}
	\quad
	\subfigure[Cross section of (a), basket call]{
		\includegraphics[width=0.45\linewidth]{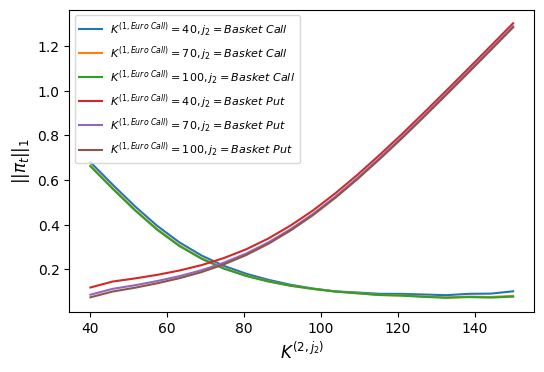}}

	\caption{$\left\Vert\bm{\pi_t}\right\Vert_1$ versus strike price (Basket option)}
    \label{GBM_basket_pi}
\end{figure}

The risky asset exposure (see Equation \eqref{chp5_l1nrom_multi}) is broken down into the  allocation on $O_t^{(1,j_1)}$ and $O_t^{(2,j_2)}$. Note that allocation on $O_t^{(2,j_2)}$ depends only on $f_t^{(22)}$ because $B_t^{(2)}$ is solely hedged by the basket option $O_t^{(2,j_2)}$.  Figure \ref{GBM_basket_pi} (a) illustrates how risky asset exposure varies with $K^{(1,j_1)}$ and $K^{(2,j_2)}$. Allocation on $O_t^{(1,j_1)}$ is scaled by the relative sensitivity $f_t^{(11)}$ (see Equation \eqref{chp5_l1nrom_multi}); hence, risky asset exposure decreases with $K^{(1,j_1)}$, and out-of-the-money $O_t^{(1,j_1)}$ is preferable regardless of the choice of $O_t^{(2,j_2)}$, for example one-asset or multi-asset options. In addition, risky asset exposure is monotonic with $K^{(2,j_2)}$ except when $K^{(1,j_1)}$ is extremely small. We plot the cross section of (a) in Figure \ref{GBM_basket_pi} (b) for illustration purposes. Given an at-the-money or out-of-the-money $O_t^{(1,j_1)}$, $\left\Vert\bm{\pi_t}\right\Vert_1$ decreases when $j_2=Basket\ Call$ and increases when $j_2=Basket\ Put$. Therefore, out-of-the-money basket options minimize risky asset exposure. Moreover, $\left\Vert\bm{\pi_t}\right\Vert_1$ is insensitive to $K^{(1,j_1)}$ because investing in basket options generally leads to a smaller allocation on $O_t^{(1,j_1)}$.

Similarly to put and call options, with the parameters listed in Table \ref{para_merton_chp5}, the optimal basket call is achieved when the strike price $K^{(2,j_2)}$ is at the upper bound, while the optimal basket put is achieved when strike price $K^{(2,j_2)}$ is at the lower bound.\\ 

Given the similarities between one-asset calls/puts and basket calls/puts, investors could be interested in the best among those choices. To answer this question, we fix the strike price of a European call $O_t^{(1,j_1)}$, letting the lower bound of a put option strike price and the upper bound of a call option strike price be proportional to the spot price:
\begin{equation*}
\begin{split}
  &[S_t^{(2)},R^{B}S_t^{(2)}]=[A^{(2,j_2)},B^{(2,j_2)}], \quad j_2\in\{Euro\ Call,Asian\ Call\}; \\ &[S_t^{(1)}+S_t^{(2)},R^{B}(S_t^{(1)}+S_t^{(2)})]=[A^{(2,j_2)},B^{(2,j_2)}], \quad j_2\in\{Basket\ Call\};\\
&[R^{A}S_t^{(2)},S_t^{(2)}]=[A^{(2,j_2)},B^{(2,j_2)}], \quad j_2\in\{Euro\ Put,Asian\ Put,Amer\ Put\};
\\&[R^{A}(S_t^{(1)}+S_t^{(2)}),S_t^{(1)}+S_t^{(2)}]=[A^{(2,j_2)},B^{(2,j_2)}], \quad j_2\in\{Basket\ Put\},
\end{split}
\end{equation*}

where $R^{A}\leq1$ and $R^{B}\geq1$.\\ 

\begin{figure}[ht]

	\centering
	\vspace{0.35cm}
	\subfigtopskip=2pt
	\subfigbottomskip=2pt
	\subfigcapskip=-5pt
	\subfigure[$K^{(1,j_1)}=40$]{
		\includegraphics[width=0.31\linewidth]{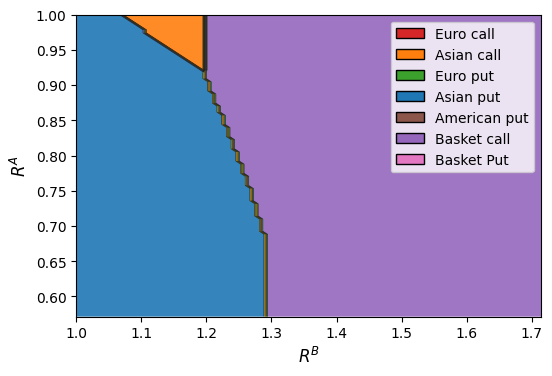}}
	\quad
	\subfigure[$K^{(1,j_1)}=50$]{
		\includegraphics[width=0.31\linewidth]{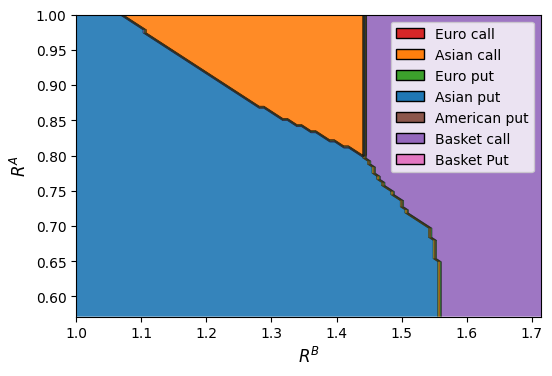}}
	\subfigure[$K^{(1,j_1)}=60$]{
		\includegraphics[width=0.31\linewidth]{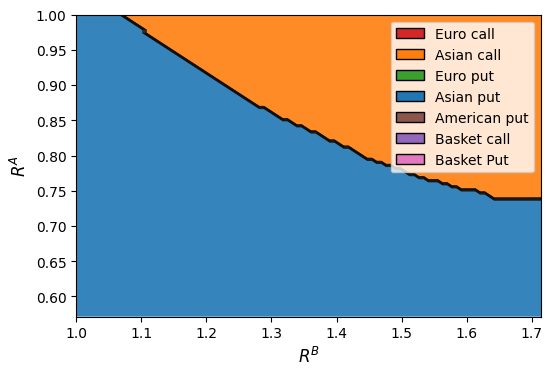}}

	\caption{Derivatives selection $O_t^{(2,j_2)}$}
    \label{GBM_basket_selection}
\end{figure}

By letting the ratio $R^{A}$ and $R^{B}$ vary, the optimal choice of $O_t^{(2,j_2)}$ is studied in Figure \ref{GBM_basket_selection}. One can observe that when $O_t^{(1,j_1)}$ is an at-the-money-option---that is, $K^{(1,j_1)}=40$---a one-asset Asian option dominates for a small $R^{B}$, while a basket call is superior to other options when $R^{B}$ is large. However, basket calls become less preferable as $K^{(1,j_1)}$ increases. As mentioned above, compared with one-asset options, investors have a smaller absolute allocation on $O_t^{(1,j_1)}$ and a larger absolute allocation on $O_t^{(2,j_2)}$ with a basket option. Furthermore, the allocation on $O_t^{(1,j_1)}$ is scaled by the relative sensitivity $f_t^{(11)}$ (see Equation \eqref{chp5_l1nrom_single} and  \eqref{chp5_l1nrom_multi}), which explains why a basket call has an advantage over one-asset options when $K^{(1,j_1)}$ is small but loses its dominant position as $K^{(1,j_1)}$ rises. The optimal choice of $O_t^{(2,j_2)}$ with other sets of parameters is demonstrated in Appendix \ref{chp5_compare_one_multi_opt}.

\section{Conclusions}

This paper reveals the benefit of using options to minimize the total risk exposure of a portfolio, while maintaining an optimal level of utility. We demonstrate that the farther out-of-the-money calls or puts are, the better choices they are,  particularly the Asian type. Given the lack of liquidity on those type of options, we explored straddle options and found that optimal choices are close to at-the-money options, which are hence likely liquid products. We also explored multi-asset derivatives and can confirm that basket options are preferable to one-asset options in terms of minimizing risk exposure.

\begin{appendices}

\section{Proofs}
\label{chp5_appendix}
\subsection{Proof of Proposition \ref{chp5_propos1}}
\label{chp5_appendixa}
We assume $V(t,W)=\frac{W_t^{1-\gamma}}{1-\gamma}\exp(h(T-t))$, which is substituted into \eqref{chp5_hjb}. Then, $h(T-t)$ satisfies:
\begin{equation}
\sup_{\bm{\pi_t}}\left\{\frac{h'(T-t)}{1-\gamma}+r+\bm{\pi_t}^T\bm{\Sigma_t\Lambda}-\frac{\gamma}{2}(\bm{\pi_t}^T\bm{\Sigma_t\Phi\Phi}^T\bm{\Sigma_t}^T\bm{\pi_t})\right\}=0,
\label{chp5_hjb1}
\end{equation}
Denote $\bm{\eta_t}=\bm{\Sigma_t}^T\bm{\pi_t}$, then problem \eqref{chp5_hjb1} can be rewritten as,
\begin{equation}
\sup_{\bm{\eta_t}}\left\{\frac{h'(T-t)}{1-\gamma}+r+\bm{\eta_t}^T\bm{\Lambda}-\frac{\gamma}{2}(\bm{\eta_t}^T\bm{\Phi\Phi}^T\bm{\eta_t})\right\}=0.
\label{chp5_hjb2}
\end{equation}
which implies the optimal strategy 
\begin{equation}
    \eta_t^*=\frac{(\bm{\Phi\Phi}^T)^{-1}\bm{\Lambda}}{\gamma}.
\end{equation}
With $\bm{\eta^*_t}=\bm{\Sigma_t}^T\bm{\pi^*_t}$, there are infinitely many choice of $\bm{\pi_t}^*$ if $n>2$. Next, we substitute $\bm{\eta^*_t}$ into \eqref{chp5_hjb2} and derive the ordinary differential equation (ODE) for $h(T-t)$:
\begin{equation}
\begin{cases}
\frac{h'(T-t)}{1-\gamma}+r+\frac{\bm{\Lambda}^T(\bm{\Phi\Phi}^T)^{-1}\bm{\Lambda}}{2\gamma}=0\\
h(T)=0.
\end{cases}\label{ode_ht}
\end{equation}
where the terminal condition results from $V(t,W)=U(W)$. The solution to \eqref{ode_ht} is
\begin{equation}
   h(T-t)=(1-\gamma)(r+\frac{\bm{\Lambda}^T(\bm{\Phi\Phi}^T)^{-1}\bm{\Lambda}}{2\gamma})(T-t).
   \end{equation}
   
\subsection{Proof of Proposition \ref{chp5_propos_twoproblem}}   
\label{appendix_twoproblem}
Let $\bm{O_{t,n}}=[O^{(1)}_t, O^{(2)}_t,..., O^{(n)}_t]^T$ with variance matrix $\Sigma_t$ of rank 2 be an optimal subset of options for problem \eqref{linear_pron}.  $\bm{\pi^*_{t,n}}$ is a strategy maximizing the expected utility if and only if $\bm{\Sigma}^T_t\bm{\pi^*_{t,n}}=\bm{\eta^*_t} $. Therefore, $\bm{O_{t,n}}$ and $\bm{\pi^*_{t,n}}$ is an optimal pair for \eqref{linear_pron} when $\bm{\pi^*_{t,n}}$ is an optimal solution for 
\begin{mini}
{\bm{\pi_t}}{\Vert\bm{\pi_t}\Vert_1}{}{}
\addConstraint{\bm{\Sigma_t}^T\bm{\pi_t}=\bm{\eta^*_t} }{}{}
\label{linear_pro_proof}
\end{mini}
According to the principle 4.5 in \cite{rardin1998optimization}, problem \eqref{linear_pro_proof} is equivalent to 
\begin{mini}
{\delta_t}{\mathbbm{1}^T\bm{\delta_t}}{}{}
\addConstraint{\bm{\hat{\Sigma}_t}^T\bm{\delta_t}=\bm{\eta^*_t }}
\addConstraint{\bm{\delta_t}\geq0}
\label{linear_pro1}
\end{mini}
where $\bm{\delta_t}=[\alpha_t^{(1)},\alpha_t^{(2)},...,\alpha_t^{(n)},\beta_t^{(1)}, \beta_t^{(2)},...,\beta_t^{(n)}]^T$ satisfies $\alpha_t^{(i)}=\frac{\vert\pi_t^{(i)}\vert+\pi_t^{(i)}}{2}$, and $\beta_t^{(i)}=\frac{\vert\pi_t^{(i)}\vert-\pi_t^{(i)}}{2}$, with
\begin{equation}
\bm{\hat{\Sigma}_t}=\left[\begin{array}{c}
\Sigma_t  \\
-\Sigma_t
\end{array}\right]=\left[\begin{array}{cc}
f^{11}_t & f^{12}_t\\
...&...\\
f^{n1}_t & f^{n2}_t\\
-f^{11}_t & -f^{12}_t \\
...&...\\
-f^{n1}_t & -f^{n2}_t 
\end{array}\right].
\end{equation}
Theorems 2.3 and 2.4 in \cite{bertsimas1997introduction} lists the necessary and sufficient conditions for the extreme point $\delta_t$, i.e.
\begin{enumerate}
    \item $\bm{\delta_t}=[\delta_t^{(1)}, \delta_t^{(2)},...,\delta_t^{(n)},\delta_t^{(n+1)}, \delta_t^{(n+2)},...,\delta_t^{(2n)}]^T$.
    \item the $\hat{q}_{th}$ and $\hat{p}_{th}$ rows in $\hat{\Sigma_t}$  are linear independent,  $\delta_t^{(i)}=0$ if $i\neq \hat{q}$ or $\hat{p}$.
    \item $\delta_t$ is feasible solution.
\end{enumerate}
Without loss of generality, we assume the $p_{th}$ and $q_{th}$ rows in $\Sigma$ are linear independent, and we consider 4 cases:
\begin{equation}
\begin{split}
\delta_t^{[1]}&=\begin{cases}
[\delta_t^{[1],(1)}, \delta_t^{[1],(2)},...,\delta_t^{[1],(n)},\delta_t^{[1],(n+1)}, \delta_t^{[1],(n+2)},...,\delta_t^{[1],(2n)}]^T\\
\delta_t^{[1],(i)}=0 \quad \textit{ if $i\neq q$ or $p$}
\end{cases}\\
\delta_t^{[2]}&=\begin{cases}
[\delta_t^{[2],(1)}, \delta_t^{[2],(2)},...,\delta_t^{[2],(n)},\delta_t^{[2],(n+1)}, \delta_t^{[2],(n+2)},...,\delta_t^{[2],(2n)}]^T\\
\delta_t^{[2],(i)}=0 \quad \textit{ if $i\neq q+n$ or $p$}
\end{cases}\\
\delta_t^{[3]}&=\begin{cases}
[\delta_t^{[3],(1)}, \delta_t^{[3],(2)},...,\delta_t^{[3],(n)},\delta_t^{[3],(n+1)}, \delta_t^{[3],(n+2)},...,\delta_t^{[3],(2n)}]^T\\
\delta_t^{[3],(i)}=0 \quad \textit{ if $i\neq q$ or $p+n$}
\end{cases}\\\delta_t^{[4]}&=\begin{cases}
[\delta_t^{[4],(1)}, \delta_t^{[4],(2)},...,\delta_t^{[4],(n)},\delta_t^{[4],(n+1)}, \delta_t^{[4],(n+2)},...,\delta_t^{[4],(2n)}]^T\\
\delta_t^{[4],(i)}=0 \quad \textit{ if $i\neq q+n$ or $p+n$}
\end{cases}
\end{split}
\end{equation}
It is clear that there is a non-negative strategy in $\delta_t^{[1]}$, $\delta_t^{[2]}$, $\delta_t^{[3]}$ and $\delta_t^{[4]}$ because the $i_{th}$ row in $\hat{\Sigma}$ is the opposite of the $(i+n)_{th}$ row, and the non-negative strategy is feasible and an extreme point. This proves the existence of an extreme point for problem \eqref{linear_pro1}.Now, theorem 2.7 in \cite{bertsimas1997introduction} guarantees that there is an optimal solution which is an extreme point for problem \eqref{linear_pro1}.\\ 
With the second necessary and sufficient conditions of the extreme point, we know that an optimal solution $\delta_t^*$ for problem \eqref{linear_pro1} has at most two non-zero elements. This would imply an optimal solution, denoted by $\bm{\pi^*_{t,n}}=[\pi_{t,n}^{(1)}, \pi_{t,n}^{(2)},...,\pi_{t,n}^{(n)}]^T$, for problem \eqref{linear_pro_proof} with at most two non-zero elements, which would also be the optimal strategy for \eqref{linear_pron}.\\ 
Without loss of generality, we assume $\pi_{t,n}^{(i)}=0$, $i\neq x,y$. $\bm{O_{t,2}}=[O^{(x)}_t, O^{(y)}_t]$ and $\bm{\pi^*_{t,2}}=[\pi_{t,n}^{(x)}, \pi_{t,n}^{(y)}]^T$ is a feasible strategy for problem \eqref{linear_pron} with $n=2$. We show that it is an optimal pair by contradiction.\\ 
If there is a feasible solution $\bm{\hat{O}_{t,n}}=[\hat{O}^{(1)}_t, \hat{O}^{(2)}_t]$ and $\bm{\hat{\pi}^*_{t,2}}=[\hat{\pi}_{t,2}^{(1)}, \hat{\pi}_{t,2}^{(2)}]^T$ such that $\Vert\bm{\hat{\pi}^*_{t,2}}\Vert_1<\Vert\bm{\pi^*_{t,2}}\Vert_1$, then $\bm{\hat{\pi}^*_{t,n}}=[\hat{\pi}_{t,2}^{(1)}, \hat{\pi}_{t,2}^{(2)},0,...,0]^T$ is a feasible strategy for \eqref{linear_pron} such that $\Vert\bm{\hat{\pi}^*_{t,n}}\Vert_1<\Vert\bm{\pi^*_{t,n}}\Vert_1$, which is contradiction to our previous conclusion. Note that $\Vert\bm{\pi^*_{t,2}}\Vert_1=\Vert\bm{\pi^*_{t,n}}\Vert_1$, so problem \eqref{linear_pron} with $n=2$ and with $n\geq 2$ have the same minimum $\ell_1$ norm of allocation.

\subsection{Proof of Proposition \ref{chp5_propos_single}}  
\label{appendix_single}
Let $O_t\in\Omega_O^{(2,1)}$ with non-singular variance matrix $\bm{\Sigma_t}$ (see \eqref{chp5_variance_matrix_single}), the optimal strategy space $\Omega_{\pi}^{O}$ contains a unique strategy, i.e. $\bm{\pi_t}=(\bm{\Sigma_t}^T)^{-1}\bm{\eta_t^{\ast}}$, and $\ell_1$ norm of $\bm{\pi_t}$ is given by
\begin{equation}
\left\Vert\bm{\pi_t}\right\Vert_1=\left\vert\frac{\eta_t^{(1)}}{f_t^{(1)}}\right\vert+\left\vert\frac{\eta_t^{(2)}}{f_t^{(2)}}\right\vert.
\label{chp5_l1nrom_single}
\end{equation}
A non-zero denominator is guaranteed by the non-singular variance matrix assumption. $O_t^{\ast}=[O_t^{{(1)},\ast},O_t^{{(2)},\ast}]^T\in\Omega_O^{(2,1)}$  with variance matrix 
\begin{equation}
\bm{\Sigma_t^*}=\left[\begin{array}{cc}
f^{(1),*}_t& 0 \\
0 & f^{(2),*}_t
\end{array}\right].
\end{equation}
For $\bm{O_t}\in\Omega_O^{(2,1)}$, if $\left\vert f^{(1)}_t\right\vert\leq \left\vert f_t^{(1),\ast}\right\vert$ and $\left\vert f^{(2)}_t\right\vert\leq \left\vert f_t^{(2),\ast}\right\vert$ hold, then it is easy to see $\left\Vert\bm{\pi_t}\right\Vert_1 \geq \left\Vert\bm{\pi_t}^{\ast}\right\Vert_1$, so $\bm{O_t^{\ast}}$ is a optimal portfolio composition.\\ 
If $\left\vert f^{(1)}_t\right\vert\leq \left\vert f_t^{(1),\ast}\right\vert$ or $\left\vert f^{(2)}_t\right\vert\leq \left\vert f_t^{(2),\ast}\right\vert$ does not hold, then there is a $\bm{O^{\ast\ast}_t}\in\Omega_O^{(2,1)}$, such that the corresponding strategy $\left\Vert \bm{\pi_t}^{\ast\ast}\right\Vert_1<\left\Vert \bm{\pi_t}^{\ast}\right\Vert_1$, hence $\bm{O_t^{\ast}}$ is not the optimal. 

We have shown that, for any $\bm{O_t}\in\Omega_O^{(2,1)}$, $\left\vert f^{(1)}_t\right\vert\leq \left\vert f_t^{(1),\ast}\right\vert$ and $\left\vert f^{(2)}_t\right\vert\leq \left\vert f_t^{(2),\ast}\right\vert$ is a sufficient and necessary condition for $\bm{O_t^{\ast}}$ to be an optimal portfolio composition for problem \eqref{linear_pro_21}. Therefore,
\begin{equation}
    \bm{O_t^{(i),*}}=\underset{O_t^{(i)}}{\arg \max }\left\vert \frac{\partial O_t^{(i)}}{\partial S_t^{(i)}}\frac{1}{O_t^{(i)}}\right\vert.
\end{equation}

\subsection{Proof of Proposition \ref{chp5_propos_putcall_exist}}
\label{appendix_putcall_exist}
$\bm{O_t^{\ast}}=[O_t^{(1),\ast},O_t^{(2),\ast}]^T$ is the optimal portfolio composition for problem \eqref{linear_pro_21} if and only if it has the largest absolute sensitivity (see Proposition \ref{chp5_propos_single}), i.e. 
\begin{equation}
    O_t^{(i),\ast}={\arg \max }\left\vert \frac{\partial O_t^{(i,j)}}{\partial S_t^{(i)}}\frac{1}{O_t^{(i,j)}}\right\vert.
\end{equation}
 For convenience, we write the absolute sensitivity as a function of strike price $KK^{(i,j)}$.
 \begin{equation*}
     h(K^{(i,j)},j)= \left\vert \frac{\partial O_t^{(i,j)}}{\partial S_t^{(i)}}\frac{1}{O_t^{(i,j)}}\right\vert
 \end{equation*}
$h(K^{(i,j)},j)$ is continuous because $O_t^{(i,j)}\in \mathbb{C}^1$ and $O_t^{(i,j)}\neq0$. According to the extreme value theorem, there is a $\hat{K}^{(i,j)}\in [A^{(i,j)},B^{(i,j)}]$ that achieves the largest $ h(K^{(i,j)},j)$, hence minimizes  $\ell_1$ norm of allocation $\left\Vert \bm{\pi_t}^*\right\Vert_1$. $O_t^{(i),\ast}$ is the optimal for problem \eqref{linear_pro_21} when the strike price $K^{(i,j)}$ is one of the $\hat{K}^{(i,j)}$  where:\\ 
$j\in\{\textit{Euro Call, Asian Call, Amer Call, Euro Put, Asian Put, Amer Put}\}$\\
This guarantees the existence of optimal composition in $\Omega_O^{(2,put\ call)}$.

\subsection{Proof of Corollary \ref{chp5_propos_limit}}  
\label{appendix_limit}
We first prove the following lemma,
\begin{lemma} \label{chp_lemma_ineq}
The inequality 
\begin{equation}
 \frac{\phi(x-c)}{N(x-c)} -c\leq \frac{\phi(x)}{N(x)}
\end{equation}
holds for $\forall x\in(-\infty,\infty), c\in(0,\infty)$, where $\phi$ and $N$ are respectively the density function and distribution function of a standard normal random variable.
\begin{proof}
Let us define the "reversed hazard rate" function:
\begin{equation*}
Y(x)=\frac{\phi(x)}{N(x)}.
\end{equation*}
We want to show%
\begin{equation*}
Y(x-c)-c\leq Y(x).
\end{equation*}
We first demonstrate $Y^{\prime }(x)\geq -1$ for all $x$.\ To see this, note:%
\begin{equation*}
 Y^{\prime }+1=\frac{\phi}{N^2}\left( \frac{\phi^{\prime }}{\phi}N-\phi+\frac{N^{2}}{\phi}%
\right) =\frac{\phi}{N^2}f.
\end{equation*}
Using $\phi^{\prime }=-x\phi$, we have $f^{\prime }=\left( 1+\frac{xN}{\phi}\right) N$%
. It is not difficult to see that $x+Y\geq 0$ (use the fact that $g=\left(
x+Y\right) N\rightarrow 0$ as $x\rightarrow -\infty $ and $g^{\prime }=N\geq
0$). Hence we know $\left( 1+\frac{xN}{\phi}\right) \geq 0$, which implies $%
f^{\prime }\geq 0$. Moreover $f(x)\geq \underset{x\rightarrow -\infty }{\lim
}f(x)=0$, therefore $Y^{\prime }+1\geq 0$.

Now we complete the proof, using $Y^{\prime }\geq -1$ for all $x$, and the
mean value theorem, we conclude "by contradiction" that
\begin{equation*}
\frac{Y(x)-Y(x-c)}{c}\geq -1
\end{equation*}%
for all $x$ and $c$, which implies%
\begin{equation*}
Y(x)\geq Y(x-c)-c.
\end{equation*}
\end{proof}
\end{lemma}

 Next, we show that absolute value of optimal allocation on an European call decreases to $0$ as $K^{(i,Euro\ Call)}\rightarrow\infty$ and absolute value of allocation on an European put increases with $K^{(i,Euro\ Put)}$ and converges to $0$ as $K^{(i,Euro\ Put)}\rightarrow0$. We first abbreviate $K^{(i,Euro\ Call)}$ to $K$. Equation \eqref{chp5_option_vol} and \eqref{chp5_euro_price} shows the sensitivity of an European call, 
\begin{equation*}
\begin{split}
\frac{\partial O^{(i,Euro\ Call)}_t}{\partial S_t^{(i)}}\frac{1}{O^{(i,Euro\ Call)}_t}  \sigma^{(i)}S_t^{(i)}&=\frac{N(d_1)\sigma^{(i)}S_t^{(i)}}{S_t^{(i)}N(d_1)-Ke^{-r(\hat{T}-t)}N(d_2)}\\&=\frac{%
\sigma^{(i)} }{1-e^{-r(\hat{T}-t)}\frac{K}{S^{(i)}_{t}}\frac{N(d_{2})}{N(d_{1})}},     
\end{split}
\end{equation*}
where
\begin{eqnarray*}
d_{1} &=&\frac{\ln {(S^{(i)}_{t}/K)}+(r+\frac{1}{2}(\sigma^{(i)} )^{2})(\hat{T}-t)}{%
\sigma^{(i)} \sqrt{\hat{T}-t}}=a\left( \ln {(S^{(i)}_{t}/K)}+r(\hat{T}-t)\right) {+}%
b\quad  \\
d_{2} &=&d_{1}-\sigma^{(i)} \sqrt{\hat{T}-t}=a\left( \ln {(S^{(i)}_{t}/K)}+r(\hat{T}%
-t)\right) {+}b-c
\end{eqnarray*}%
with  $c=\sigma^{(i)} \sqrt{\hat{T}-t}>0$, $b=\frac{1}{2}c$ and $a=\frac{1}{c}$. Let us rewrite the
sensitivity as follows:%
\begin{equation*}
\frac{\partial O^{(i,Euro\ Call)}_t}{\partial S_t^{(i)}}\frac{1}{O^{(i,Euro\ Call)}_t}  \sigma^{(i)}S_t^{(i)}=\frac{%
\sigma^{(i)}}{1-e^{\frac{b-x}{a}}G(x)}
\end{equation*}
where $x=a\left( \ln {(S_{t}/K)}+r(\hat{T}-t)\right) {+}b\,$\ and $G(x)=%
\frac{N(d_{2})}{N(d_{1})}=\frac{N(x-c)}{N(x)}$.\\ 
We would like to show $%
H(x)=e^{\frac{b-x}{a}}G(x)$ is increasing in $K$, hence decreasing in $x$. Its first derivative leads to:%
\begin{eqnarray*}
H^{\prime }(x) &=&-\frac{e^{\frac{b-x}{a}}}{a}G(x)+e^{\frac{b-x}{a}%
}G^{\prime }(x)\\
&=&-\frac{e^{\frac{b-x}{a}}}{a}\frac{N(x-c)}{N(x)}+e^{\frac{b-x%
}{a}}\frac{n(x-c)N(x)-N(x-c)n(x)}{N^{2}(x)} \\
&=&-\frac{e^{\frac{b-x}{a}}}{a}\frac{N(x-c)}{N(x)}+e^{\frac{b-x}{a}}\left(
\frac{n(x-c)}{N(x-c)}-\frac{n(x)}{N(x)}\right) \frac{N(x-c)}{N(x)} \\
&=&e^{\frac{b-x}{a}}\left[ -\frac{1}{a}+\left( \frac{n(x-c)}{N(x-c)}-\frac{%
n(x)}{N(x)}\right) \right] \frac{N(x-c)}{N(x)} \\
&=&e^{\frac{b-x}{a}}\left[ -c+\left( \frac{n(x-c)}{N(x-c)}-\frac{n(x)}{N(x)}%
\right) \right] \frac{N(x-c)}{N(x)}
\end{eqnarray*}
It's easy to see $H^{\prime }(x)<0$ with Lemma \ref{chp_lemma_ineq}. The sensitivity of an European call is positive and increases with $K$.\\

As $K\rightarrow \infty$, $O^{(i, Euro\ call)}_t\rightarrow0$, so $Ke^{-r(\hat{T}-t)}N(d_2)\rightarrow0$. Furthermore,
\begin{eqnarray*}
H(x)&=&\frac{Ke^{-r(\hat{T}-t)}N(d_2)}{S_t^{(i)}N(d_1)} \xrightarrow{\text{L'Hôpital's rule}} \frac{e^{-r(\hat{T}-t)}N(d_2)-\phi(d_1)\frac{S_t^{(i)}}{K\sigma^{(i)}\sqrt{\hat{T}-t}}}{-\frac{S_t^{(i)}\phi(d_1)}{K\sigma^{(i)}\sqrt{\hat{T}-t}}}\\
&\xrightarrow{\text{L'Hôpital's rule}}&\frac{-\phi(d_1)\frac{S_t^{(i)}}{(K)^2\sigma^{(i)}\sqrt{\hat{T}-t}}-\phi(d_1)\frac{d_1S_t^{(i)}}{(K\sigma^{(i)}\sqrt{\hat{T}-t} )^2} +\frac{\phi(d_1)S_t^{(i)}}{(K)^2\sigma^{(i)}\sqrt{\hat{T}-t}}}{-\phi(d_1) \frac{S_t^{(i)}d_1}{(K\sigma^{(i)}\sqrt{\hat{T}-t} )^2}+\frac{S_t^{(i)}\phi(d_1)}{(K)^2\sigma^{(i)}\sqrt{\hat{T}-t}}}
\\&=&\frac{-S_t^{(i)}\sigma^{(i)}\sqrt{\hat{T}-t}-S_t^{(i)}d_1+S_t^{(i)}\sigma^{(i)}\sqrt{\hat{T}-t}}{-S_t^{(i)}d_1+S_t^{(i)}\sigma^{(i)}\sqrt{\hat{T}-t}}\longrightarrow 1.
\label{Nd1/K_call}
\end{eqnarray*}
Hence,
\begin{equation*}
\frac{\partial O^{(i,Euro\ Call)}_t}{\partial S_t^{(i)}}\frac{1}{O^{(i,Euro\ Call)}_t}  \sigma^{(i)}S_t^{(i)}=\frac{%
\sigma^{(i)}}{1-H(x)}\longrightarrow\infty.
\end{equation*}
Moreover, the absolute value of allocation on $O_t^{(i,Euro\ Call)}$ is decreasing with $K$ and 
\begin{equation}
    \left\vert\pi^{(i,Euro\ Call)}_t\right\vert=\frac{ \left\vert\eta_t^{(i)}\right\vert}{ \left\vert\frac{\partial O_t^{(i,Euro\ Call)}}{\partial S_t^{(i)}}\frac{1}{O_t^{(i,Euro\ Call)}}  \sigma^{(i)}S_t^{(i)}\right\vert}\longrightarrow 0 \quad \textit{ as } K\longrightarrow\infty.
\end{equation}

The proof for European put follows similarly.

\subsection{Proof of Proposition \ref{chp5_propos_straddle}}
\label{appendix_straddle}
Suppose for any $\bm{O_{t}}\in\Omega _{O}^{(2,straddle)}$, $O_{t}^{(i,j)}$ satisfies these four conditions:

\begin{multicols}{2}
\begin{enumerate}
    \item $\frac{\partial O_t^{(i,j)}}{\partial S_t^{(i)}}=0 \iff K^{(i,j)}=A^{(i,j)} $.
    \item $\left\vert\frac{\partial O_t^{(i)}}{\partial S_t^{(i)}}\right\vert$ has an upper bound.
    \item $O_t^{(i,j)}\in(0,\infty)$ and $O_{t}^{(i,j)}\rightarrow\infty$ as $K^{(i,j)}\rightarrow\infty$, where $j\in\{Euro\ Strad, Asian\ Strad\}$.
    \item $O_t^{(i,j)}\left(S^{(i)},K^{(i,j)}\right)\in \mathbb{C}^1$.
\end{enumerate}
\end{multicols}
All four assumptions are not restrictive in a Black-Scholes setting. Proposition \ref{chp5_propos_single} illustrates that $\bm{O_t^{\ast}}=[O_t^{(1),\ast},O_t^{(2),\ast}]^T$  is the optimal portfolio composition for problem \eqref{linear_pro_21} if it has the largest absolute sensitivity, i.e. 
\begin{equation}
    O_t^{(i),\ast}=\underset{O_t^{(i,j)}}{\arg \max }\left\vert \frac{\partial O_t^{(i,j)}}{\partial S_t^{(i)}}\frac{1}{O_t^{(i,j)}}\right\vert.
\end{equation}
For convenience, we write the absolute sensitivity as a function of strike price $K^{(i,j)}$.
 \begin{equation*}
     h(K^{(i,j)},j)= \left\vert \frac{\partial O_t^{(i,j)}}{\partial S_t^{(i)}}\frac{1}{O_t^{(i,j)}}\right\vert
 \end{equation*}
 With the four assumptions above, it's easy to see,
\begin{equation}
\begin{cases}
h(0,j)\in(0,\infty)& j\in\{Euro\ Strad, Asian\ Strad, Amer\ Strad\} \\
h(B^{(i,j)},j)\in(0,\infty)& j=Amer\ Strad\\
h(K^{(i,j)},j)\rightarrow0 & \textit{as }K^{(i,j)} \uparrow\downarrow A^{(i,j)}, j\in\{Euro\ Strad, Asian\ Strad, Amer\ Strad\} \\
h(K^{(i,j)},j)\rightarrow0 & \textit{as }K^{(i,j)} \rightarrow\infty, j\in\{Euro, Asian\} .
\end{cases}
\end{equation}
When $K^{(i,j)}\in[0,A^{(i,j)})$ and $j\in\{Euro\ Strad, Asian\ Strad, Amer\ Strad\}$, $h(K^{(i,j)},j)$ is continuous because\\ $O_t^{(i,j)}\left(S^{(i)},K^{(i,j)}\right)\in \mathbb{C}^1$. Besides, there is a $Z$ such that $h(K^{(i,j)},j)<h(0,j)$ when $K^{(i,j)}\in(Z,A^{(i,j)})$. According to the extreme value theorem, there is a $K^{(l,i,j)}$ such that $h$ attains the maximum in $[0,Z]$, hence $h(K^{(l,i,j)},j)\geq h(0,j)$. $K^{(l,i,j)}$ is proven to be the maximum point for $h(K^{(i,j)},j)$ in $[0,A^{(i,j)})$.\\
Let $M$ be a real number in $(A^{(i,j)},\infty)$ where $j\in\{Euro\ Strad, Asian\ Strad\}$, it's obvious that $h(M,j)>0$. There is a $Z>0$ such that $h(K^{(i,j)},j)<h(M,j)$ when $K^{(i,j)}\in(A^{(i,j)},A^{(i,j)}+\frac{1}{Z})\cup(Z,\infty)$. According to the extreme value theorem, there is a $K^{(r,i,j)}$ such that $h$ attains the maximum in $[A^{(i,j)}+\frac{1}{Z},Z]$, i.e. $h(K^{(r,i,j)},j)>h(M,j)$. Then, $K^{(r,i,j)}$ is the maximum point for $h(K^{(i,j)},j)$ in $(A^{(i,j)},\infty)$.\\ 
As for American straddle ($j=Amer\ Strad$), $h(B^{(i,j)},j)>0$. There is a $Z>0$ such that $h(K^{(i,j)},j)<h(B^{(i,j)},j)$ when $K^{(i,j)}\in(A^{(i,j)},A^{(i,j)}+\frac{1}{Z})$. And there is a $K^{(r,i,j)}$ such that $h$ attains the maximum in $[A^{(i,j)}+\frac{1}{Z},B^{(i,j)}]$, hence $h(K^{(r,i,j)},j)\geq h(B^{(i,j)},j)$. $h(K^{(r,i,j)},j)$ is the maximum point on the right branch $[A^{(i,j)},B^{(i,j)}]$. $O_t^{(i),\ast}$ is the optimal for the problem \eqref{linear_pro_21} when the strike price $K$ is either $K^{(l,i,j)}$ or $K^{(r,i,j)}$ where $j\in\{Euro\ Strad, Asian\ Strad, Amer\ Strad\}$. The existence of the optimal composition in a straddle subset is proven.
 
\subsection{Proof of Proposition \ref{chp5_propos_multi}}  
\label{appendix_multi}
 Let $\bm{O_t}\in\Omega_O^{(2,multi\ asset)}$ with non-singular variance matrix $\bm{\Sigma_t}$ (see \eqref{chp5_variance_matrix_multi}), the optimal strategy space $\Omega_{\pi}^{O}$ contains a unique strategy, i.e. $\bm{\pi_t}=(\bm{\Sigma_t}^T)^{-1}\bm{\eta_t^{\ast}}$. The allocation and its $\ell_1$ norm can be written as
\begin{equation}
\begin{split}
\pi^{(1,j)}&=\frac{1}{ f^{(11)}_t}(\eta_t^{(1)}-\frac{f^{(21)}_t}{f^{(22)}_t}\eta_t^{(2)})\quad\quad \pi^{(2,j)}=\frac{\eta_t^{(2)}}{f^{(22)}_t}
\\
\left\Vert\bm{\pi_t}\right\Vert_1&=\frac{1}{\left\vert f^{(11)}_t\right\vert}\left\vert\eta_t^{(1)}-\frac{f^{(21)}_t}{f^{(22)}_t}\eta_t^{(2)}\right\vert+\frac{\left\vert\eta_t^{(2)}\right\vert}{\left\vert f^{(22)}_t\right\vert}. 
\end{split}
\label{chp5_l1nrom_multi}
\end{equation}
If $\bm{O_t^{\ast}}=[O_t^{{(1),\ast}},O_t^{{(2),\ast}}]^T\in\Omega_O^{(2,multi\ asset)}$  achieves minimum $\ell_1$ norm of allocation and 
\begin{equation}
    O_t^{(1),\ast}\neq\underset{O_t^{(1,j_1)}}{\arg \max }\left\vert \frac{\partial O_t^{(1,j_1)}}{\partial S_t^{(1)}}\frac{1}{O_t^{(1,j_1)}}\right\vert.
\end{equation}
Then there is a $\bm{O_t^{\ast\ast}}=[O_t^{{(1),\ast\ast}},O_t^{{(2),\ast\ast}}]^T$, such that
\begin{equation}
    \left\vert \frac{\partial O_t^{(1),\ast}}{\partial S_t^{(1)}}\frac{1}{O_t^{(1),\ast}}\right\vert<\left\vert \frac{\partial O_t^{(1),\ast\ast}}{\partial S_t^{(1)}}\frac{1}{O_t^{(1),\ast\ast}}\right\vert.
\end{equation}
Therefore, let $O_t^{{(2),\ast\ast}}=O_t^{{(2)},\ast}$, this implies $\left\vert f^{(11),\ast}_t\right\vert<\left\vert f^{(11),\ast\ast}_t\right\vert$, $f^{(21),\ast}_t=f^{(21),\ast\ast}_t$ and $f^{(22),\ast}_t=f^{(22),\ast\ast}_t$. Equation  \eqref{chp5_l1nrom_multi} indicates  $\left\Vert\bm{\pi_t}^{\ast}\right\Vert_1>\left\Vert\bm{\pi_t}^{\ast\ast}\right\Vert_1$, which proves by contradiction that
\begin{equation}
    O_t^{(1),\ast}=\underset{O_t^{(1,j_1)}}{\arg \max }\left\vert \frac{\partial O_t^{(1,j_1)}}{\partial S_t^{(1)}}\frac{1}{O_t^{(1,j_1)}}\right\vert.
\end{equation}
is a necessary condition for $\bm{O_t^{\ast}}$ to be the optimal portfolio composition for problem \eqref{linear_pro_21} within $\Omega_O^{(2,multi\ asset)}$.

 \subsection{Proof of Proposition \ref{chp5_propos_callbasket_exist}}
\label{appendix_callbasket_exist}
Recall $f_t^{(ik)}$ in the variance matrix $\Sigma_t$ can be written as 
\begin{equation}
 f^{(ik)}  = \frac{\partial O_t^{(i,j_i)}}{\partial S_t^{(k)}}\frac{S_t^{(k)}\sigma^{(k)}}{O_t^{(i,j_i)}}.
\end{equation}
According to the non-singular variance matrix assumption, $f_t^{(11)},f_t^{(22)}\neq0$. In addition, $O_t^{(1,j_1)}$ is an European call option, hence $f_t^{(11)}$ is continuous with respect to $K^{(1,Euro\ Call)}$ on  $[A^{(1,Euro\ Call)},B^{(1,Euro\ Call)}]$. $f_t^{(21)}$ and $f_t^{(22)}$ are  continuous with respect to $K^{(2,j_2)}$ on  $[A^{(2,j_2)},B^{(2,j_2)}]$ because $O_t^{(2,j_2)}\in \mathbb{C}^1$.\\
$\left\Vert\bm{\pi_t}\right\Vert_1$ (see \eqref{chp5_l1nrom_multi}) is continuous on the closed set $[A^{(1,Euro\ Call)},B^{(1,Euro\ Call)}]\times[A^{(2,j_2)},B^{(2,j_2)}]$, so there is a portfolio with strike price $[\hat{K}^{(1,Euro\ Call)},\hat{K}^{(2,j_2)}]^T$ that achieves minimum risky asset exposure with any $j_2$. $\bm{O_t^{\ast}}$ is the optimal for problem \eqref{linear_pro_21} within $\Omega_O^{(2,call\ basket)}$ when the strike price is in $[\hat{K}^{(1,Euro\ Call)},\hat{K}^{(2,j_2)}]^T$, where $j_2\in\{\textit{Basket Call, Basket Put}\}$.

\subsection{Comparison between the one-asset option and multi-asset option subsets} \label{chp5_compare_one_multi_opt}
\begin{figure}[ht]
	\centering
	\vspace{0.35cm}
	\subfigtopskip=2pt
	\subfigbottomskip=2pt
	\subfigcapskip=-5pt
	\subfigure[$K^{(1,j_1)}=40$]{
		\includegraphics[width=0.31\linewidth]{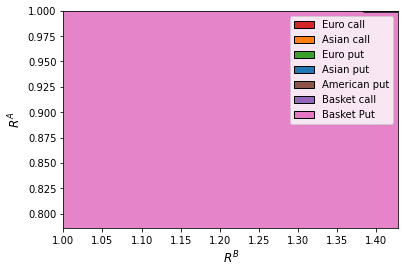}}
	\quad
	\subfigure[$K^{(1,j_1)}=50$]{
		\includegraphics[width=0.31\linewidth]{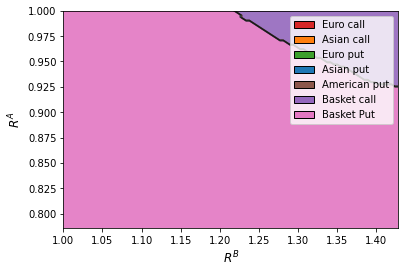}}
	\subfigure[$K^{(1,j_1)}=60$]{
		\includegraphics[width=0.31\linewidth]{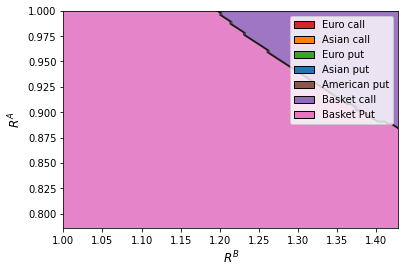}}

	\caption{Derivatives selection $O_t^{(2,j_2)}$ ($\rho=-0.4$)}
    \label{GBM_basket_selection_case2}
\end{figure}

In this section, we exhibit the optimal choice of $O_t^{(2,j_2)}$ given different set of parameters. In contrast to the two positive correlated underlying assets considered in Section \ref{chp5_5}, i.e. $\rho=0.4$ (see Table \ref{para_merton_chp5}), we let correlation $\rho=-0.4$ while all other parameters remain unchanged, here a similar derivatives selection is conducted.\\ 
Results are presented in Figure \ref{GBM_basket_selection_case2}. Unlike the case of two positive correlated underlying assets (see Figure \ref{GBM_basket_selection}),  one-asset option is no longer  preferable in minimizing risky asset exposure while basket put becomes competitive. Especially when $O_t^{(1,j_1)}$ is an at-the-money European call, i.e. $K^{(1,j_1)}=40$, a basket put is superior to other options, i.e. larger area. 

\begin{figure}[ht]

	\centering
	\vspace{0.35cm}
	\subfigtopskip=2pt
	\subfigbottomskip=2pt
	\subfigcapskip=-5pt
	\subfigure[$K^{(1,j_1)}=40$]{
		\includegraphics[width=0.31\linewidth]{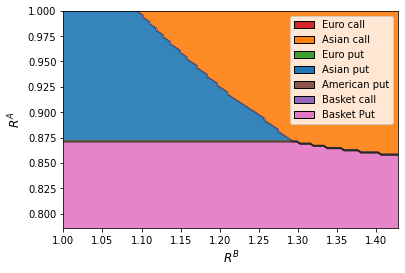}}
	\quad
	\subfigure[$K^{(1,j_1)}=50$]{
		\includegraphics[width=0.31\linewidth]{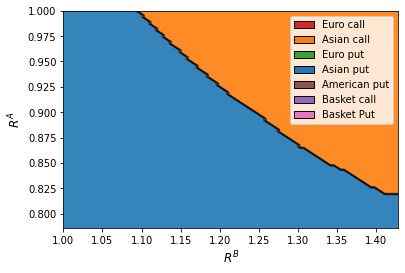}}
	\subfigure[$K^{(1,j_1)}=60$]{
		\includegraphics[width=0.31\linewidth]{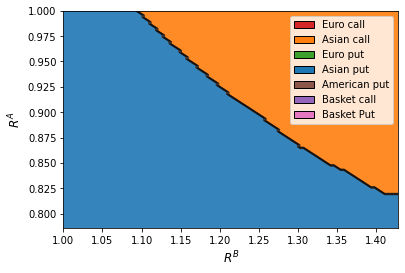}}

	\caption{Derivatives selection $O_t^{(2,j_2)}$ (exchange assets' parameter)}
    \label{GBM_basket_selection_case3}
\end{figure}

Next, we consider the case when the parameters of the two underlying assets are exchanged, i.e. $\lambda^{(1)}=0.6$, $\lambda^{(2)}= 0.52 $, $\sigma^{(1)}=0.2$, $\sigma^{(2)}=0.13$ while all other parameters are given in Table \ref{para_merton_chp5}, the optimal choice of $O_t^{(2,j_2)}$ is shown in Figure \ref{GBM_basket_selection_case3}.  Compared with Figure \ref{GBM_basket_selection}, basket put instead of basket call is selected in the largest region. Furthermore, the one-asset Asian option is preferable when $K^{(1,j_1)}$ is lregardless of the underlying assets' parameters. 

\end{appendices}


\bibliography{bib}

\begin{thebibliography}{14}
\providecommand{\natexlab}[1]{#1}
\providecommand{\url}[1]{{#1}}
\providecommand{\urlprefix}{URL }
\providecommand{\doi}[1]{\url{https://doi.org/#1}}
\providecommand{\eprint}[2][]{\url{#2}}
 \bibcommenthead

\bibitem[{Bertsimas and Tsitsiklis(1997)}]{bertsimas1997introduction}
Bertsimas D, Tsitsiklis JN (1997) Introduction to linear optimization, vol~6.
  Athena Scientific Belmont, MA

\bibitem[{Bjerksund and Stensland(1993)}]{bjerksund1993closed}
Bjerksund P, Stensland G (1993) Closed-form approximation of american options.
  Scandinavian Journal of Management 9:S87--S99

\bibitem[{Black and Scholes(1973)}]{black1973pricing}
Black F, Scholes M (1973) The pricing of options and corporate liabilities.
  Journal of political economy 81(3):637--654

\bibitem[{Chen and Liu(2014)}]{chen2014american}
Chen N, Liu Y (2014) American option sensitivities estimation via a generalized
  infinitesimal perturbation analysis approach. Operations Research
  62(3):616--632

\bibitem[{Goldman et~al(1979)Goldman, Sosin, and Gatto}]{goldman1979path}
Goldman MB, Sosin HB, Gatto MA (1979) Path dependent options:" buy at the low,
  sell at the high". The Journal of Finance 34(5):1111--1127

\bibitem[{Haugh and Lo(2001)}]{haugh2001asset}
Haugh MB, Lo AW (2001) Asset allocation and derivatives. Quantitative Finance
  1:45--72

\bibitem[{Kemna and Vorst(1990)}]{kemna1990pricing}
Kemna AG, Vorst AC (1990) A pricing method for options based on average asset
  values. Journal of Banking \& Finance 14(1):113--129

\bibitem[{Kraft(2003)}]{kraft2003elasticity}
Kraft H (2003) Elasticity approach to portfolio optimization. Mathematical
  Methods of Operations Research 58(1):159--182

\bibitem[{Liu and Pan(2003)}]{liu2003dynamic}
Liu J, Pan J (2003) Dynamic derivative strategies. Journal of Financial
  Economics 69(3):401--430

\bibitem[{Maverick(2020)}]{maverick_2020}
Maverick J (2020) How big is the derivatives market?
  \urlprefix\url{https://www.investopedia.com/ask/answers/052715/how-big-derivatives-market.asp}

\bibitem[{Merton(1969)}]{merton1969lifetime}
Merton RC (1969) Lifetime portfolio selection under uncertainty: The
  continuous-time case. The review of Economics and Statistics pp 247--257

\bibitem[{Merton(1973)}]{merton1973theory}
Merton RC (1973) Theory of rational option pricing. The Bell Journal of
  economics and management science pp 141--183

\bibitem[{Mi and Xu(2020)}]{mi2020optimal}
Mi H, Xu L (2020) Optimal investment with derivatives and pricing in an
  incomplete market. Journal of Computational and Applied Mathematics
  368:112,522

\bibitem[{Rardin and Rardin(1998)}]{rardin1998optimization}
Rardin RL, Rardin RL (1998) Optimization in operations research, vol 166.
  Prentice Hall Upper Saddle River, NJ

\end{thebibliography}


\end{document}